\documentclass{aastex61}
\usepackage{graphicx}	% Including figure files
\usepackage{amsmath}	% Advanced maths commands
\usepackage{amssymb}	% Extra maths symbols

\usepackage{longtable} % for the long table 

\usepackage{threeparttablex}

\usepackage{threeparttable} 
\usepackage[section]{placeins} % to put each fig. and table in its place (very useful)
\received{2017}
\revised{}
\accepted{\today}
\submitjournal{ApJ}

\shorttitle{The $^{14}$N/$^{15}$N ratio in OMC-2 FIR4}
\shortauthors{Kahane et al.}
\begin{document}

\title{First measurement of the $^{14}$N/$^{15}$N ratio in the
  analogue of the Sun progenitor OMC-2 FIR4}

\correspondingauthor{Claudine Kahane}
\email{claudine.kahane@univ-grenoble-alpes.fr}

\author{Claudine Kahane}
\affiliation{Univ. Grenoble Alpes, CNRS, IPAG, 122 rue de la Piscine, 38000 Grenoble, France}

\author{Ali Jaber Al-Edhari}
\affiliation{Univ. Grenoble Alpes, CNRS, IPAG, 122 rue de la Piscine, 38000 Grenoble, France}
\affiliation{University of Al-Muthanna, College of Science, Physics Department, Al-Muthanna, Iraq}

\author{Cecilia Ceccarelli}
\affiliation{Univ. Grenoble Alpes, CNRS, IPAG, 122 rue de la Piscine, 38000 Grenoble, France}

\author{Ana L\'opez-Sepulcre}
\affiliation{Institut de Radioastronomie Millimetrique, 300 rue de la Piscine, 38406, Saint-Martin d'Heres, France }

\author{Francesco Fontani}
\affiliation{INAF-Osservatorio Astrofisico di Arcetri, Largo E. Fermi 5, I-50125, Florence, Italy}

\author{Mihkel Kama}
\affiliation{Institute of Astronomy, University of Cambridge, Madingley Road, Cambridge CB3 0HA, UK}

%%%%%%%%%%%%%%%%%%%%%%%%%%%%%%%%%%%%%%%%%%%

\begin{abstract}
  We present a complete census of the $^{14}$N/$^{15}$N isotopic ratio
  in the most abundant N-bearing molecules towards {the cold envelope} of the protocluster
  OMC-2 FIR4, the best known Sun progenitor.  To this scope, we
  analysed the unbiased spectral survey obtained with the IRAM-30m
  telescope at 3mm, 2mm and 1mm. We detected several lines of CN, HCN, HNC,
  HC$_3$N, N$_2$H$^+$, and their respective $^{13}$C and $^{15}$N
  isotopologues. The lines relative fluxes are compatible with LTE
  conditions and moderate line opacities have been corrected via a Population
  Diagram method or theoretical relative intensity ratios of the
  hyperfine structures. The five species lead to very similar
  $^{14}$N/$^{15}$N isotopic ratios, without any systematic difference
  between amine and nitrile bearing species as previously found in
  other protostellar sources. The weighted average of the
  $^{14}$N/$^{15}$N isotopic ratio is $270\pm 30$. This $^{14}$N/$^{15}$N value
  is remarkably consistent with  {the [250$-$350] range measured for the local galactic ratio but significantly differs from the ratio measured in comets (around 140). High-angular resolution observations are needed to examine whether this discrepancy is maintained at smaller scales.}
  In addition, using
  the CN, HCN and HC$_3$N lines, we derived a $^{12}$C/$^{13}$C
  isotopic ratio of {$50\pm 5$}. 
\end{abstract}


%
%%%%%%%%%%%%%%%%%%%%%%%%%%%%%%%%%%%%%%%%%%%%%%%%%%%
\section{Introduction} \label{sec:intro}

The Solar System is the result of a long and complex process, several
aspects of which still remain a mistery. One of these is the so-called
"anomalous" $^{14}$N/$^{15}$N value in the objects of the Solar
System~\citep{Caselli:2012,Hily-Blant:2013}. Based on the solar wind
particles~{\citep{Marty:2012}}, the Solar Nebula value is
$441\pm 6$. However, $^{14}$N/$^{15}$N is  { 272} in the Earth
atmosphere~{\citep{Marty:2012}}, around 140 in
comets~\citep{Manfroid:2009,Mumma:2011,Shinnaka:2014,Rousselot:2014},
and between 5 and 300 in
meteorites~\citep{Busemann:2006,Bonal:2010,Aleon:2010}. The Solar
System primitive objects as well as the terrestrial atmosphere are,
hence, enriched of $^{15}$N with respect to the presumed initial
value. It has been long known that, similarly to the $^{15}$N
enrichment, the D/H ratio in terrestrial water is about ten times
larger than in the Solar Nebula and this is very likely due to the
conditions in the earliest phases of the Solar System (see e.g. the
reviews by  ~\cite{Ceccarelli:2014a} and ~\cite{Cleeves:2015}).
The reason for the $^{15}$N enrichment has, thus, been searched for in
the chemical evolution of matter during the first steps of the Solar
System formation
(e.g.~\cite{Terzieva:2000,Rodgers:2008,Wirstrom:2012,Hily-Blant:2013}). 

Several observations in Solar-like star forming regions have been
reported in {the} literature. In prestellar cores, $^{14}$N/$^{15}$N  varies
between 70 and {more than 1000} (\cite{Ikeda:2002,Gerin:2009,Milam:2012}; {\cite{Daniel:2013}}; \cite{Hily-Blant:2013,Hily-Blant:2017,Bizzocchi:2013,Taniguchi:2017}), in Solar-like Class 0 protostars between {150} and {600} \citep{Gerin:2009, Wampfler:2014}, and 80$-$160 in protoplanetary disks~\citep{Guzman:2015,Guzman:2017}.

{Whereas the $^{14}$N/$^{15}$N values reported in the literature
  for prestellar cores, protostars, disks and comets have been derived
  from the observations of half a dozen of different species (CN, HCN,
  HNC, NH$_3$, N$_2$H$^+$, cyanopolyynes), it should be noted that,
  for each of these objects, only a few species were used each time.}

Nonetheless, one has to consider that, rather than to the Solar Nebula
$^{14}$N/$^{15}$N value, these measurements should be compared to the
nowadays local interstellar $^{14}$N/$^{15}$N ratio of $\sim {300}$,
which results from cosmic evolution in the solar neighbourhood
(\cite{Romano:2017,Hily-Blant:2017} and references therein) and is, apparently by
coincidence, very close to the terrestrial atmosphere value.

In order to understand the origin of the Solar System $^{15}$N
enrichment, we need to measure it in objects that are as most as
possible similar to the Sun progenitor.  The so far known best
analogue of the Sun progenitor is represented by the source OMC-2 FIR4,
in the Orion Molecular Complex at a distance of 420
pc~\citep{Menten:2007,Hirota:2007}, north of the famous KL
object. Several recent observations show that FIR4 is a young
protocluster containing several protostars, some of which will
eventually become
Suns~\citep{Shimajiri:2008,Lopez-Sepulcre:2013,Furlan:2014}. In
addition, OMC-2 FIR4 shows signs of the presence of one or more sources
of energetic $\geq 10$ MeV particles, the dose of which is similar to
that measured in meteoritic material \citep{Ceccarelli:2014,Fontani:2017}. 
In this article, we report the first measure of the
$^{14}$N/$^{15}$N ratio in OMC-2 FIR4, using different molecules:
HC$_3$N, HCN, HNC, CN and N$_2$H$^+$.

\section{Observations} \label{obs} We carried out an unbiased spectral
survey of OMC-2 FIR4 (PI: Ana L\'opez-Sepulcre) in the 1,2 and 3mm
bands with the IRAM\,30\,m telescope. The 3\,mm (80.5$-$116.0\,GHz) and
2\,mm (129.2$-$158.8\,GHz) bands were observed between 31 Aug. and 5
Sep. 2011, and on 24 Jun. 2013. The 1\,mm (202.5$-$266.0\,GHz) range was
observed on 10$-$12 Mar. 2012 and on 7 Feb. 2014.  The Eight MIxer
Receiver (EMIR) has been used, connected to the 195~kHz resolution
Fourier Transform Spectrometer (FTS) units. The observations were
conducted in wobbler switch mode, with a throw of 120$''$. Pointing
and focus measurements were performed regularly.  
{The telescope Half Power Beam Width (HPBW)
  is 21$-$30.6$''$, 15.5$-$19$''$ and 9$-$12$''$  in the 1,2 and 3mm bands respectively.}
The package CLASS90
of the GILDAS software
collection\footnote{http://www.iram.fr/IRAMFR/GILDAS/} was used to
reduce the data. The uncertainties of calibration are estimated to be
lower than 10\% at 3\,mm and 20\% at 2 mm and 1mm. After subtraction of the
continuum emission via first-order polynomial fitting, a final
spectrum was obtained by stitching the spectra from each scan and
frequency setting. The intensity was converted from antenna
temperature ($T_\mathrm{ant}^\ast$) to main beam temperature
($T_\mathrm{mb}$) using the beam efficiencies provided at the IRAM web
site. { The typical rms noise, expressed in $T_\mathrm{mb}$ unit, is 4$-$7 mK in the 3mm band, 8$-$10mK in the 2mm band, and 15$-$25mK in the 1mm band.} 
\section{Results} \label{result} In order to derive the
$^{14}$N/$^{15}$N ratio in OMC-2 FIR4, we have looked for the
$^{15}$N-bearing substitutes of all the abundant N-bearing species
present in the survey. We have clearly detected and identified the
$^{15}$N-isotopologues of five species: HC$_3$N, HCN, HNC, CN,
N$_2$H$^+$, most with more than one line, and tentatively detected one
line of H$^{13}$C$^{15}$N. In addition, we have also included in our
analysis the $^{13}$C-bearing isotopes of HC$_3$N, HCN, HNC and CN.  A
representative sample of these lines is shown in
Fig.~\ref{fig:sample-spectra} {and all the observed lines are plotted in the Appendix in Figs.~\ref{fig:obs_HC3N} to \ref{fig:obs_13C15N}.}

%%%%%%%%%% spectra lines %%%%%%%%
\begin{figure}[htb]
  \centering
 \plotone{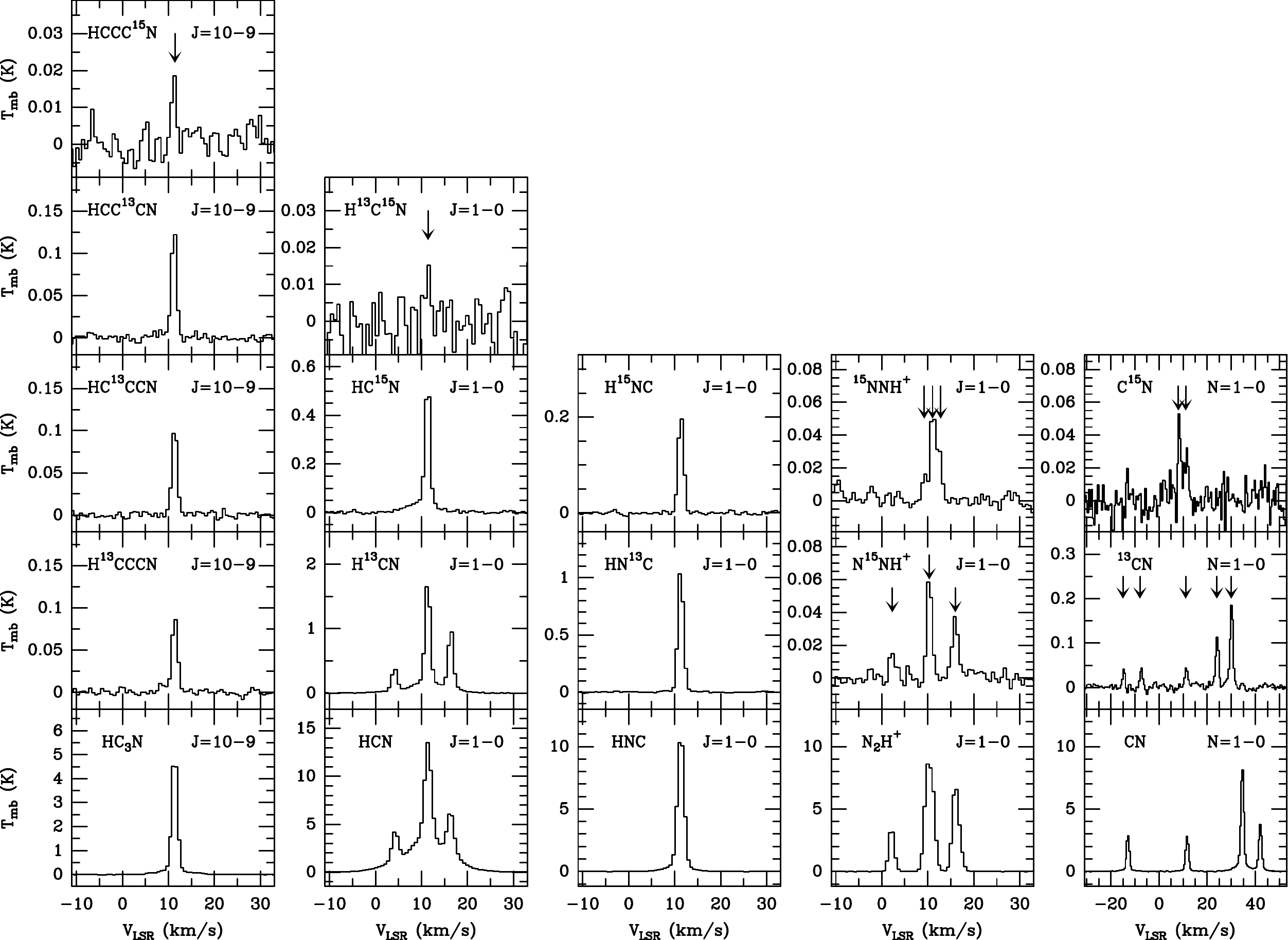}
  \caption{Representative sample of the observed spectra for the
    HC$_3$N, HCN, HNC, N$_2$H$^+$ and CN families. The temperature
    scale is main beam temperature. Weak lines are indicated by an
    arrow.}
  \label{fig:sample-spectra}
\end{figure}
%%%%%%%%%%%%%%%%%%%%%%%%%%%%%%%%%%

{ The lines analysis and modeling presented here make use of several tools of the CASSIS package.}\footnote{(CASSIS:Centre d'Analyse Scientifique de Spectres Instrumentaux et Synthétiques): is a line analysis and modeling
  software developed by IRAP\-UPS/CNRS ({\it
    http://cassis.irap.omp.eu})} Gaussian fits have been used to derive the lines
integrated intensities (called « fluxes » in the following) and kinematical properties. All lines are well fitted with
narrow { G}aussian components showing low dispersions in central
velocities and linewidths (V$_{LSR}$ = 11.3 (0.1) km.s$^{-1}$, FWHM = 1.4
(0.2) km.s$^{-1}$). In addition, the strongest lines 
also show a slightly displaced broad component (V$_{LSR}$ = 10.8 (0.3)
km.s$^{-1}$, FWHM = 6.4 (0.4) km.s$^{-1}$ from { G}aussian fits), which is not detected
in the $^{15}$N-bearing species, except in HC$^{15}$N. We
have, thus, focussed our work on the narrow component. 
{  Figs.~\ref{fig:obs_HC3N} to ~\ref{fig:obs_13C15N} of the Appendix show the Gaussian profiles superimposed to the observed lines.}
The fluxes reported in Table~\ref{tab:HC3N} {to Table~\ref{tab:CN}} correspond to the narrow
{ G}aussian components only and the 1 sigma error bars include the fit
and the calibration uncertainties.

{For HCN, HNC and N$_2$H$^+$, our survey covers only the $(J=1-0)$ transitions of the $^{15}$N-bearing isotopologues. Thus, for these species, to derive of the  $^{14}$N/$^{15}$N ratio we used the "flux ratio method" applied to the $(J=1-0)$ transitions of the observed isotopologues. It leads to reliable abundance ratios provided that (i) the $(1-0)$ transitions of the various  isotopologues correspond to the same excitation temperature, (ii) the lines are not significantly affected by or can be corrected for opacity effects, (iii) the emission size is the same for the various isotopologues.}

{For HC$_3$N and CN, since more than one rotational transition is observed for the $^{15}$N-bearing isotopologues,  we could perform a Local Thermal Equilibrium (LTE) modeling to derive the $^{14}$N/$^{15}$N abundance ratios, as discussed for each species below.} 

Whenever possible, we have
obtained direct $^{14}$N/$^{15}$N measurements. In two cases, HCN and
HNC, we have obtained indirect $^{14}$N/$^{15}$N derivations from the
less abundant isotopologues H$^{13}$CN and HN$^{13}$C, assuming a
$^{12}$C/$^{13}$C ratio.

{ In all cases, the isotopic ratios that we derive are « beam-averaged » values, at the scale of the largest HPBW of our observations ($\sim 30"$)}.

\subsection{HC$_3$N}
To rely on a coherent set of lines, likely to sample the same gas,
{ we have restricted our analysis to the narrow HC$_3$N emission, because
broad emission from the $^{15}$N and the $^{13}$C isotopes of
HC$_3$N is  not detected in our survey.}

\begin{figure}
  \centering
\plotone{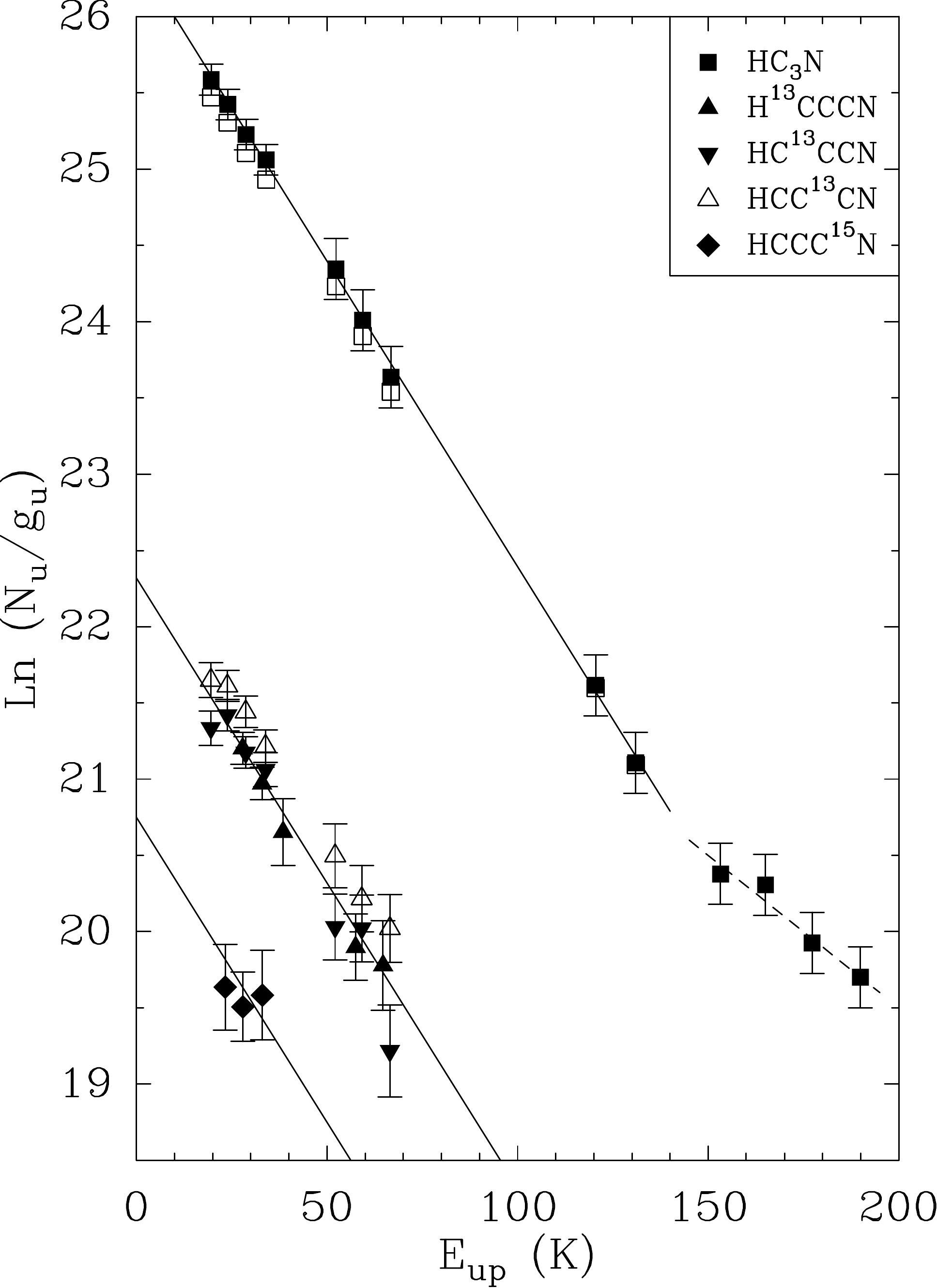}
 \epsscale{0.5}
  \caption{Population Diagrams for HC$_3$N and its isotopologues. No beam dilution correction has been applied. The HC$_3$N data points are shown as black squares when corrected for opacity effects and    as empty squares without the opacity correction. The three solid lines correspond to a rotational temperature of 25 K;{ the dashed line corresponds to an excitation temperature of 50 K.} }
\label{fig:pd_tot}
\end{figure}

%%%%%%%%%%%%%% Table 1 : HC3N + iso %%%%%%%%%%%
\begin{table*}[h]
  \centering
   \caption{Integrated intensities from Gaussian fits for HC$_3$N and its isotopologues.}
  \begin{tabular}{|c|c|c|c|c|c|c|c|}
    \hline
\multicolumn{4}{|c|}{HC$_3$N}    & H$^{13}$CCCN   & HC$^{13}$CCN    & HCC$^{13}$CN    &  HCCC$^{15}$N     \\ 
\hline
Transition & Frequency$^{(1)}$ & E$_{up}^{(1)}$   & $\int$T$_{mb}$dv         &  $\int$T$_{mb}$dv        &  $\int$T$_{mb}$dv        &  $\int$T$_{mb}$dv &  $\int$T$_{mb}$dv\\
  $(J'-J)$   & [MHz]     & [K]         &  [K.km.s$^{-1}$]       & [K.km.s$^{-1}$]       & [K.km.s$^{-1}$]        & [K.km.s$^{-1}$]        &  [K.km.s$^{-1}$]          \\
   \hline 
    9$-$8    & 81881.5 & 19.6        & 7.1(0.7)        &    -           & 0.11(0.01)      & 0.16(0.02)      &    -                \\
    10$-$9    & 90979.0 & 24.0        & 7.4(0.7)        & 0.14(0.01)     & 0.16(0.02)      & 0.18(0.02)      & 0.024(0.007)       \\
    11$-$10   & 100076.4 & 28.8        & 7.3(0.7)        & 0.15(0.02)     & 0.16(0.02)      & 0.19(0.02)      & 0.028(0.006)       \\
    12$-$11   & 109173.6 & 34.0        & 7.4(0.7)        & 0.14(0.01)     & 0.17(0.02)      & 0.18(0.02)      & 0.030(0.010)        \\
    13$-$12   & 114615.0$^{(2)}$
 &  38.5$^{(2)}$          &   -             & 0.11(0.01)     &    -            &   -             &   -                 \\
    15$-$14   & 136464.4
 & 52.4        & 5.7(1.1)        &  -             & 0.10(0.02)      & 0.13(0.03)      &   -                 \\
    16$-$15   & 145561.0  & 59.4        & 4.7(0.9)        & 0.08(0.02)     & 0.10(0.02)      & 0.12(0.02)      &   -                 \\
    17$-$16   & 154657.3 & 66.8        & 3.7(0.7)        & 0.08(0.02)     & 0.06(0.01)      & 0.13(0.03)      &   -                 \\
24$-$23     & 218324.7  & 130.9   & 0.6(0.1)   &   - &   - &   - &   -    \\
26$-$25     & 236512.8 & 153.2   & 0.36(0.07)  &   - &   - &   - &   -       \\
27$-$26     & 245606.3 & 165.0   & 0.37(0.07)  &   - &   - &   - &   -       \\
29$-$28     & 263792.3 & 189.9   & 0.22(0.04)  &   - &   - &   - &   -       \\ 
\hline    
\end{tabular}
\label{tab:HC3N}
\begin{tablenotes}
\item[1]{\footnotesize The integrated intensities correspond to the narrow { G}aussian components only and the 1 $\sigma$ error bars include the fit and the calibration uncertainties.}  
\item[2]{\footnotesize (1) We report in this table only the main isotopologue frequencies and upper level energies, except for the (13$-$12) transition, which falls outside the observed frequency range.}
\item[2]{\footnotesize (2) These values correspond to the H$^{13}$CCCN isotopologue, the only one for which the (13$-$12) transition is covered by our observations.}
\end{tablenotes}
\end{table*}
%%%%%%%%%%%%%%%%%%%%%%%%%%%%%%%%%%
To derive the column densities of the five species, we have performed a Population Diagram analysis of the lines reported in
Table~\ref{tab:HC3N}, which corrects iteratively for the opacity
effects. 
{Large scale maps obtained recently with the IRAM 30m telescope (Jaber Al-Edhari et al., in prep.)
show that the cyanopolyyne emission is extended. Thus, no beam
dilution correction has been applied.}
The results of the analysis are shown in Fig.~\ref{fig:pd_tot}.

%%%%%%%%%% Population Diagram of HC3N %%%%%%%%
%%%%%%%%%%%%%%%%%%%%%%%%%%%%%%%%%%

{ The HC$_3$N lines diagram suggests the existence of two
  components responsible for the narrow emission}.
After correction of their (moderate) opacity (0.20 to 0.25), the
HC$_3$N lines { with upper level energy E$_{up}$ lower than 130 K 
 are compatible with an LTE excitation at} a rotational temperature of 25 K,  
in agreement with the values derived
from the $^{13}$C isotopes, and a HC$_3$N column density of
3.3$\times$10$^{13}$ cm$^{-2}$. Assuming the same excitation
temperature for the HCCC$^{15}$N lines, we obtain a column density of
1.2$\times$10$^{11}$ cm$^{-2}$ and a direct determination of the
$^{14}$N/$^{15}$N abundance ratio of $275\pm 65$. For the three
$^{13}$C-bearing isotopologues of HC$_3$N, we derive, in the same way,
$^{12}$C/$^{13}$C abundance ratios of $57\pm 7$, $59\pm 11$ and
$46\pm 6$.
{We have checked that an LTE model based on these parameters is
perfectly coherent with the non detection of the isotopologues lines
at  E$_{up}$ $>$ 70 K. }

{The HC$_3$N lines with  E$_{up}$ $>$150 K
suggest the existence of a warmer component but its analysis is out
of the scope of this article. An extended and detailed modeling of
  the cyanopolyynes emission in OMC-2 FIR4, relying on a more complete
  set of data and including the broad emission, is presented in
  another study (Jaber Al-Edhari et al., in prep.).}

\subsection{HCN}

{The lines of HCN and of its  $^{13}$C- and $^{15}$N- bearing isotopologues show  a narrow and
a broad emission. We have fitted them
by Gaussian profiles (see Table~\ref{tab:HCN} and  Fig.~\ref{fig:obs_HCN}).}

%%%%%%%%%%%%%% Table 2 : HCN + iso %%%%%%%%%%%
\begin{table*}[h]
  \centering
   \caption{Integrated intensities from Gaussian fits for HCN and its isotopologues.}
  \begin{tabular}{|c|c|c|c|c|c|c|c|}
    \hline
\multicolumn{5}{|c|}{HCN}    & H$^{13}$CN  &  HC$^{15}$N    & H$^{13}$C$^{15}$N \\ 
\hline
Transition & Frequency$^{(2)}$ & E$_{up}^{(2)}$  & Component$^{(3)}$    & $\int$T$_{mb}$dv$^{(4)}$          &  $\int$T$_{mb}$dv$^{(4)}$        &  $\int$T$_{mb}$dv$^{(4)}$        &  $\int$T$_{mb}$dv$^{(4)}$   \\
    $(J'_{F'}-J_{F})$ or $(J'-J)$  $^{(1)}$    & [MHz] & [K]     &    &  [K.km.s$^{-1}$]       & [K.km.s$^{-1}$]       & [K.km.s$^{-1}$]        &  [K.km.s$^{-1}$]          \\
   \hline 
$1_1-0_1$ & 88630.4 & 4.3 & N & 8.4(0.9) & 1.3(0.1) & - & -\\
$1_2-0_1$ or $1-0$ & 88631.9 & 4.3 & N & 13(2) & 2.2(0.2) & 0.70(0.07) & 0.020(0.008) \\
               &  &   & B & 50 (5) & 2.0 (0.2) & 0.34 (0.03) & -  \\
$1_0-0_1$ & 88633.9 & 4.3 & N & 5.2(0.5) & 0.44(0.06) & - & -\\
$3-2$ &265886.4 & 25.2 & N & 24(1) & 1.3(0.3) & - & -\\
 & &   & B & 85(18) & 7.2(1.5) & - & -\\

\hline    
\end{tabular}
\label{tab:HCN}
\begin{tablenotes}
\item[1]{\footnotesize (1) For the $^{15}$N-bearing isotopologues and the broad Gaussian component, the reported transition is the $(J=1-0)$ one.}
\item[2]{\footnotesize (2) We report in this table only the main isotopologue frequencies and upper level energies.}
\item[3]{\footnotesize (3) Narrow (N) and broad (B) Gaussian components fitted to the observed spectra, with no attempt to distinguish the hyperfine components in the broad emission.}
\item[4]{\footnotesize (4) The 1 $\sigma$ error bars include the fit and the calibration uncertainties.}  
\end{tablenotes}
\end{table*}
%%%%%%%%%%%%%%%%%%%%%%%%%%%%%%%%%%
{\subsubsection{Narrow emission}}

The $(J=1-0)$ spectra of HCN and H$^{13}$CN split into three hyperfine
components, which provides a measure of the { line} opacity when
comparing the relative intensities of the hyperfine components. Under
LTE optically thin conditions, this ratio is 1:3:5.  
Relative to the
weakest component, the observed flux ratios obtained for HCN and
H$^{13}$CN are respectively 1:(1.6$\pm$0.2):(2.6$\pm$0.5) and 1:(2.9$\pm$0.5):(5.1$\pm$0.9) (see Table~\ref{tab:check}). It suggests that
the H$^{13}$CN {hyperfine components have the same excitation temperature and} are optically thin, whereas the HCN lines suffer from significant opacity or/and anomalous excitation
effects, which will prevent from a direct determination of the
$^{14}$N/$^{15}$N ratio. Neglecting the weak differences of line
frequencies, the double isotopic ratio
$^{13}$C$^{14}$N/$^{12}$C$^{15}$N is then simply equal to the ratio of
the total fluxes, obtained by adding the hyperfine component
contributions: $5.6\pm 0.8$. This may lead to an indirect determination
of the $^{14}$N/$^{15}$N ratio, assuming a $^{12}$C/$^{13}$C ratio. In
addition, the H$^{13}$C$^{15}$N $(1-0)$ line being tentatively detected
(at 2.5 sigmas, see Fig.~\ref{fig:sample-spectra} and  Fig.~\ref{fig:obs_HCN}), the comparison with the
H$^{13}$C$^{14}$N $(1-0)$ total flux provides a direct measurement of
the $^{14}$N/$^{15}$N ratio, equal to $200\pm 85$, whereas the
comparison with the HC$^{15}$N $(1-0)$ flux gives a $^{12}$C/$^{13}$C
ratio of $36\pm 15$. It should be noted that these ratios may be somewhat underestimated, the
H$^{13}$C$^{15}$N $(1-0)$ line being only tentatively detected, which
tends to an overestimation of its flux.

{In addition, the Population Diagram built with the $J=(1-0)$ and $(2-1)$ fluxes of H$^{13}$CN leads to an
excitation temperature of 6 K, if the emission is assumed to be more extended than the largest beam (30$''$), a reasonable assumption for such cold gas.}

{\subsubsection{Broad emission}}

{The broad emission is particularly evident in the $(J=2-1)$ spectra of HCN
and H$^{13}$CN, but it is also detected in the $(J=1-0)$ emission, even for HC$^{15}$N.
We have estimated the corresponding $^{13}$C$^{14}$N/$^{12}$C$^{15}$N
double isotopic ratio from the $(J=1-0)$ line flux ratio. This value, $6.0\pm 1.5$, shows a
quite large uncertainty but is fully compatible with the ratio derived
from the narrow emission. 
A Population Diagram applied to the H$^{13}$CN $(J=1-0)$  and $(J=2-1)$ fluxes, with 
no beam dilution,  leads to an excitation temperature of 22 K, which suggests that this broad emission could come from a warmer component than the narrow one.}

\subsection{HNC}
{The broad emission, which is marginally visible in the HNC $(1-0)$
and HN$^{13}$C $(2-1)$ spectra only, has not been included in
our analysis (see Table~\ref{tab:HNC-N2H+} and Fig.~\ref{fig:obs_HNC-N2H+}).}
The HNC hyperfine structure is too narrow to be spectrally resolved in
our data. However, {a  Population Diagram analysis applied to} the $(J=1-0)$ and $(2-1)$ lines of HN$^{13}$C, {assuming extended emission,} indicates  moderate opacities
($\sim$0.2) and leads to an excitation temperature of 7$-$8 K. {Such a low excitation temperature appears coherent with the assumption of extended emission.}

As the HNC line is certainly more severely affected by opacity effect, a direct
derivation of the $^{14}$N/$^{15}$N ratio is not possible.
From the flux ratio of $^{13}$C- and $^{15}$N- bearing species, the double
isotopic ratio $^{13}$C$^{14}$N/$^{12}$C$^{15}$N equal to
$5.4\pm 0.8$, becomes $6.0\pm 0.8$ when the opacity correction is applied to the HN$^{13}$C line flux.

%%%%%%%%%%%%%% Table 3 : HNC and N2H+ + iso %%%%%%%%%%%
\begin{table*}[h]
  \centering
   \caption{Integrated intensities from Gaussian fits for HNC, N$_2$H$^+$ and their isotopologues.}
  \begin{tabular}{|c|c|c|c|c|c|c|c|}
\hline 
\multicolumn{4}{|c|}{HNC}    & \multicolumn{2}{c|}{HN$^{13}$CN}  &  \multicolumn{2}{c|}{H$^{15}$NC}    \\ 
    \hline
    Transition & E$_{up}$   & Frequency & $\int$T$_{mb}$dv         & Frequency &  $\int$T$_{mb}$dv    & Frequency     &  $\int$T$_{mb}$dv \\
 $(J'-J)$  & [K]  & [MHz]            &  [K.km.s$^{-1}$]    & [MHz]   & [K.km.s$^{-1}$]     & [MHz]   & [K.km.s$^{-1}$]        \\
\hline          
$1-0$ & 4.4 & 90663.6 & 21(2) & 87090.9 & 1.7(0.2) & 88865.7 &  0.31(0.01) \\
$3-2$ & 25.1 & & - & 261263.3 & 1.09(0.04) & & - \\
\hline    
\multicolumn{4}{|c|}{N$_2$H$^+$}    & \multicolumn{2}{c|}{$^{15}$NNH$^+$}  &  \multicolumn{2}{c|}{N$^{15}$NH$^+$}    \\ 
    \hline
    Transition & E$_{up}$   & Frequency & $\int$T$_{mb}$dv         & Frequency$^{(1)}$ &  $\int$T$_{mb}$dv    & Frequency$^{(1)}$     &  $\int$T$_{mb}$dv \\
$(J'_{F'}-J_{F})$  & [K]  & [MHz]            &  [K.km.s$^{-1}$]    & [MHz]   & [K.km.s$^{-1}$]     & [MHz]   & [K.km.s$^{-1}$]        \\

\hline
$1_1-0_1$ & 4.5 &93171.9 & 13.1(1.5) &90263.5 & 0.08(0.01) &91204.3 & 0.059(0.008) \\
$1_2-0_1$ & 4.5 &93173.7 & 21.3(2.5) &90263.9 & 0.036(0.005) & 91206.0 & 0.09(0.02) \\
$1_0-0_1$ & 4.5 &93176.1 & 5.3(0.5) &90264.5 & 0.015(0.004) &91208.5 & 0.025(0.006) \\
\hline 
\end{tabular}
\label{tab:HNC-N2H+}
\begin{tablenotes}
\item[1]{\footnotesize The integrated intensities correspond to the narrow { G}aussian components only and the 1 $\sigma$ error bars include the fit and the calibration uncertainties.}  
\item[2]{\footnotesize (1) Frequencies from ~\cite{Dore:2009}.}
\end{tablenotes}
\end{table*}

\subsection{N$_2$H$^+$}
The main isotope and each of the $^{15}$N-bearing substitutes of
N$_2$H$^+$ show three hyperfine components,
with relative intensities of 1:3:5 if LTE optically thin emission
applies (see Table~\ref{tab:check}). The frequencies of the hyperfine lines of $^{15}$NNH$^+$ and
N$^{15}$NH$^+$ have been presented by~\cite{Dore:2009}. The observed
hyperfine flux ratios are 
1:(2.5$\pm$0.4):(4.1$\pm$0.6) for the main isotopologue,
1:(2.4$\pm$0.7):(5.0$\pm$1.5) for $^{15}$NNH$^+$ and
1:(2.4$\pm$0.6):(3.8$\pm$1.2) for N$^{15}$NH$^+$.  We conclude that,
as expected, the emission of the $^{15}$N-bearing species is optically
thin and that the lines opacity is very moderate for the main isotopologue
components. Assuming that the weakest line of N$_2$H$^+$ $(1-0)$ is
optically thin, we can estimate the “opacity corrected fluxes” of the
two others. With such a correction, the $^{14}$N/$^{15}$N ratios derived from the total fluxes are $320\pm 60$
from $^{15}$NNH$^+$ and $240\pm 50$ from N$^{15}$NH$^+$.

\subsection{CN}
The CN family members present extremely rich rotational spectra,
combining fine and hyperfine structure interactions {and our survey
covers both the $(N=1-0)$ and $(N=2-1)$ transitions for the three
isotopologues (see Figs.~\ref{fig:obs_CN} and \ref{fig:obs_13C15N}).}

{In addition, the main isotopologue shows a broad emission, more visible on the $(N=2-1)$ transitions than on the $(N=1-0)$ ones.}

{Most of the} CN $(N=1-0)$ {and} $(N=2-1)$ hyperfine components reported in the CDMS and JPL databases are easily detected. 
We have compared the observed flux ratios with the theoretical ratios
(proportional to the g$_{up}$ . A$_{ij}$ ratios, the slight frequency
differences being neglected). The results (see Table~\ref{tab:check}) suggest that the hyperfine components
follow an intensity distribution very close to LTE and that the line
opacities are moderate. 

{The same analysis shows that the $^{13}$CN and  C$^{15}$N 
$(N=1-0)$ and $(N=2-1)$ which are clearly detected and do
not suffer from blending follow an intensity distribution very close to LTE and that the lines
are optically thin.} 

{We have thus performed a simultaneous LTE modeling of the $(N=1-0)$ and $(N=2-1)$ transitions for the three isotopologues. For $^{13}$CN and C$^{15}$N we have assumed that the emission comes from a single extended component whose kinematical properties are derived from the gaussian fits (V$_{LSR}$ = 11.4  km.s$^{-1}$, FWHM = 1.3 km.s$^{-1}$). For CN, to account for the broad emission, we have added a second component (V$_{LSR}$ = 11.1  km.s$^{-1}$, FWHM = 6.9 km.s$^{-1}$). The free parameters of our modelling, performed  with a Markov Chain Monte-Carlo (MCMC) minimization to obtain the best fit to the lines, were the excitation temperature $T_{ex}$ and the column densities of the three isotopologues for the narrow emission component, the source size, the CN column density and the  excitation temperature for the broad CN emission component.}

{For the first component, the best fit was obtained with the following parameters: $T_{ex}$ = $8\pm 1$ K, $N$(CN) = (3.5$\pm$0.5)$\times$10$^{14}$ cm$^{-2}$, $N$($^{13}$CN) = (8$\pm$1)$\times$10$^{12}$ cm$^{-2}$, $N$(C$^{15}$N) = (1.3$\pm$0.2)$\times$10$^{12}$ cm$^{-2}$. It corresponds to the following isotopic ratios: $^{13}$C$^{14}$N/$^{12}$C$^{15}$N = $6.2\pm 1.3$,
$^{12}$C/$^{13}$C = $44\pm 8$ and $^{14}$N/$^{15}$N = $270\pm 60$.}

{For the broad component, there are three free parameters and the best fit solution is degenerate. However, the excitation temperature depends only weakly on the assumed size and is between 50 and 60 K.}

{The calculated profiles are superimposed to the observed ones in Figs.~\ref{fig:obs_CN} and Fig.~\ref{fig:obs_13C15N}.}

\startlongtable
\begin{deluxetable}{ccccccc}
\tablecaption{Integrated intensities from Gaussian fits for CN and its isotopologues \tablenotemark{} \label{tab:CN}}

    \tablehead{
\colhead{Species} & \colhead{Transition$^{(1)}$} & \colhead{Frequency} & \colhead{E$_{up}$}   & \colhead{V$_{LSR}$} & \colhead{FWHM} & \colhead{$\int$T$_{mb}$dv} \\ 
\colhead{}  & $N_{J'F'1F'}-N_{JF1F}$ or $N'_{J'F'}-N_{JF}$  & \colhead{[MHz]}  & \colhead{[K]} & \colhead{[km.s$^{-1}$]} & \colhead{[km.s$^{-1}$]}   & \colhead{[K.km.s$^{-1}$]}  
}
\startdata
% & line & freq & Eup & VLSR & err & FWHM & err & Flux & err
CN & 1$_{0~1/2~1/2}$ $-$ 0$_{0~1/2~1/2}$ & 113123.37 & 5.43 & 11.6(0.3) & 1.4(0.3) & 0.9(0.1) \\
 & 1$_{0~1/2~1/2}$ $-$ 0$_{0~1/2~3/2}$ & 113144.19 & 5.43 &  11.7(0.3) & 1.5(0.3) & 4.9(0.5) \\
 & 1$_{0~1/2~3/2}$ $-$ 0$_{0~1/2~1/2}$ & 113170.54 & 5.43 &  11.7(0.3) & 1.5(0.3) & 5.3(0.5) \\
 & 1$_{0~1/2~3/2}$ $-$ 0$_{0~1/2~3/2}$ & 113191.33 & 5.43 & 11.7(0.3) & 1.5(0.3) & 5.6(0.6) \\
 & 1$_{0~3/2~3/2}$ $-$ 0$_{0~1/2~1/2}$ & 113488.14 & 5.45 & 11.7(0.3) & 1.5(0.3) & 6.0(0.7) \\
 & 1$_{0~3/2~5/2}$ $-$ 0$_{0~1/2~3/2}$ & 113490.99 & 5.45 & 11.7(0.3) & 1.6(0.3) & 14.0(1.6) \\
 & 1$_{0~3/2~1/2}$ $-$ 0$_{0~1/2~1/2}$ & 113499.64 & 5.45 & 11.6(0.3) & 1.4(0.3) & 4.3(0.4) \\
 & 1$_{0~1/2~1/2}$ $-$ 0$_{0~1/2~3/2}$ & 113508.93 & 5.45 & 11.7(0.3) & 1.4(0.3) & 4.5(0.5) \\
 & 1$_{0~3/2~1/2}$ $-$ 0$_{0~1/2~3/2}$ & 113520.42 & 5.45 & 11.6(0.3) & 1.3(0.3) & 0.8(0.1) \\
 & 2$_{0~3/2~1/2}$ $-$ 1$_{0~3/2~1/2}$ & 226287.43 & 16.31 & 11.3(0.1) & 1.1(0.1) & 0.5(0.1) \\
 & 2$_{0~3/2~1/2}$ $-$ 1$_{0~3/2~3/2}$ & 226298.92 & 16.31 & 11.0(0.2) & 1.6(0.2) & 0.6(0.2) \\
 & 2$_{0~3/2~3/2}$ $-$ 1$_{0~3/2~1/2}$ & 226303.08 & 16.31 & 11.3(0.1) & 1.2(0.1) & 0.5(0.1) \\
 & 2$_{0~3/2~3/2}$ $-$ 1$_{0~3/2~3/2}$ & 226314.54 & 16.31 & 11.2(0.1) & 1.1(0.1) & 1.0(0.2) \\
 & 2$_{0~3/2~3/2}$ $-$ 1$_{0~1/2~5/2}$ & 226332.54 & 16.31 & 11.3(0.1) & 1.1(0.1) & 0.5(0.1) \\
 & 2$_{0~3/2~5/2}$ $-$ 1$_{0~3/2~3/2}$ & 226341.93 & 16.31 & 11.3(0.1) & 1.1(0.1) & 0.7(0.1) \\
 & 2$_{0~3/2~5/2}$ $-$ 1$_{0~3/2~5/2}$ & 226359.87 & 16.31 & 11.3(0.1) & 1.2(0.1) & 2.6(0.6) \\
 & 2$_{0~3/2~1/2}$ $-$ 1$_{0~1/2~3/2}$ & 226616.56 & 16.31 & 11.3(0.1) & 1.1(0.1) & 0.5(0.1) \\
 & 2$_{0~3/2~3/2}$ $-$ 1$_{0~1/2~3/2}$ & 226632.19 & 16.31 & 11.3(0.1) & 1.3(0.1) & 3.2(0.7) \\
 & 2$_{0~3/2~5/2}$ $-$ 1$_{0~1/2~3/2}$ & 226659.58 & 16.31 & 11.4(0.1) & 1.6(0.1) & 8.3(1.7) \\
 & 2$_{0~3/2~1/2}$ $-$ 1$_{0~1/2~1/2}$ & 226663.70 & 16.31 & 11.4(0.1) & 1.4(0.1) & 3.2(0.7) \\
 & 2$_{0~3/2~3/2}$ $-$ 1$_{0~1/2~1/2}$ & 226679.38 & 16.31 & 11.4(0.1) & 1.3(0.1) & 3.6(0.8) \\
 & 2$_{0~5/2~5/2}$ $-$ 1$_{0~3/2~5/2}$ & 226892.12 & 16.34 & 11.3(0.1) & 1.2(0.1) & 3.2(0.7) \\
 & 2$_{0~5/2~3/2}$ $-$ 1$_{0~3/2~5/2}$ & 226905.38 & 16.34 & 11.3(0.2) & 0.9(0.2) & 0.10(0.03) \\
\hline
$^{13}$CN & 1$_{1/2~0~1}$ $-$ 0$_{1/2~1~1}$ & 108412.86 & 5.23 & 11.0(0.4) & 1.1(0.4) & 0.03(0.01) \\
 & 1$_{1/2~0~1}$ $-$ 0$_{1/2~1~2}$ & 108426.89 & 5.23 & 11.4(0.3) & 1.3(0.3) & 0.07(0.01) \\
 & 1$_{3/2~1~0}$ $-$ 0$_{1/2~0~1}$ & 108631.12 & 5.21 & 11.7(0.4) & 1.3(0.4) & 0.04(0.02) \\
 & 1$_{3/2~1~1}$ $-$ 0$_{1/2~0~1}$ & 108636.92 & 5.21 & 11.3(0.3) & 1.4(0.3) & 0.11(0.02) \\
 & 1$_{3/2~1~2}$ $-$ 0$_{1/2~0~1}$ & 108651.30 & 5.21 & 11.5(0.3) & 1.4(0.3) & 0.18(0.03) \\
 & 1$_{1/2~1~2}$ $-$ 0$_{1/2~1~2}$ & 108657.65 & 5.24 & 11.4(0.3) & 1.3(0.3) & 0.13(0.02) \\
 & 1$_{3/2~2~3}$ $-$ 0$_{1/2~1~2}$ & 108780.20 & 5.25 & 11.4(0.3) & 1.5(0.3) & 0.27(0.05) \\
 & 1$_{3/2~2~2}$ $-$ 0$_{1/2~1~1}$ & 108782.37 & 5.25 & 11.4(0.3) & 1.5(0.3) & 0.16(0.03) \\
 & 1$_{3/2~2~1}$ $-$ 0$_{1/2~1~0}$ & 108786.98 & 5.25 & 11.3(0.3) & 1.5(0.3) & 0.06(0.01) \\
 & 1$_{3/2~2~1}$ $-$ 0$_{1/2~1~1}$ & 108793.75 & 5.25 & 11.3(0.3) & 1.4(0.3) & 0.06(0.01) \\
 & 1$_{3/2~2~2}$ $-$ 0$_{1/2~1~2}$ & 108796.40 & 5.25 & 11.4(0.4) & 1.1(0.4) & 0.05(0.02) \\
 & 2$_{5/2~3~3}$ $-$ 1$_{3/2~2~2}$ & 217467.15 & 15.69 & 11.3(0.1) & 1.3(0.1) & 0.36(0.08) \\
 & 2$_{5/2~3~2}$ $-$ 1$_{3/2~2~1}$ & 217469.15 & 15.69 & 11.2(0.1) & 1.8(0.1) & 0.23(0.05) \\
\hline
C$^{15}$N & 1$_{1/2~1}$ $-$ 0$_{1/2~1}$ & 109689.61 & 5.27 & 11.1(0.4) & 2.1(0.4) & 0.05(0.02) \\
 & 1$_{3/2~1}$ $-$ 0$_{1/2~0}$ & 110023.54 & 5.28 & 11.5(0.3) & 1.3(0.3) & 0.05(0.02) \\
 & 1$_{3/2~2}$ $-$ 0$_{1/2~1}$ & 110024.59 & 5.28 & 11.4(0.4) & 1.3(0.4) & 0.08(0.02) \\
 & 2$_{3/2~2}$ $-$ 1$_{1/2~1}$ & 219722.49 & 15.81 & 11.3(0.2) & 1.3(0.2) & 0.08(0.02) \\
 & 2$_{5/2~2}$ $-$ 1$_{3/2~1}$ & 219934.04 & 15.84 & 11.4(0.1) & 1.7(0.1) & 0.10(0.03) \\
\enddata
\begin{tablenotes}
\item[1]{\footnotesize {(1) For CN and $^{13}$CN, according to the CDMS convention, the quantum numbers are N, J, F$_1$, F with F$_1$ = J + I$_1$, F = F$_1$ + I$_2$ where I$_1$ is the $^{12}$C or $^{13}$C nuclear spin and I$_2$ that of $^{14}$N. For C$^{15}$N, the quantum numbers are N, J, F with J = N + S and F = J + I, where S and I are respectively the electronic spin and the nuclear spin of $^{15}$N.}}
\item[2]{\footnotesize (2) {The kinematical parameters and} the integrated intensities correspond to the narrow { G}aussian components only and the 1 $\sigma$ error bars include the fit and the calibration uncertainties.}
\end{tablenotes}
\end{deluxetable}

%%%%%%%%%%%%%%%%%%%%%%%%%%%%%%%%%%%%%%%%%%%%%%%%%%%%%%%%%

\section{Discussion and conclusions} \label{discussion} 

\subsection{The $^{14}$N/$^{15}$N ratio towards OMC-2 FIR4}
We have reported here a complete census of the $^{14}$N/$^{15}$N ratio
in the most abundant N-bearing species towards the protocluster
OMC-2 FIR4. The five $^{14}$N/$^{15}$N ratios derived directly from the
line fluxes (HC$_3$N, CN, N$_2$H$^+$, H$^{13}$CN) appear very similar.
Their weighted average is {$260\pm 30$}. Two indirect $^{14}$N/$^{15}$N
ratio derivations, obtained with HCN and HNC, can be made from the
double isotopic ratios $^{13}$C$^{14}$N / $^{12}$C$^{15}$N. Discussing
the $^{12}$C/$^{13}$C isotopic ratio is outside of the scope of this
study. However, it can just be noticed that one of the
$^{13}$C-bearing isotopologues of HC$_3$N seems to show systematically
stronger lines than the two others (see Fig.~\ref{fig:pd_tot}) but this
requires complementary observations to be confirmed and discussed.
Although far from complete in terms of $^{12}$C/$^{13}$C measurements,
our data allow to derive five direct estimates of this isotopic
ratio and its weighted average value is {$50\pm 5$}. The resulting
indirect $^{14}$N/$^{15}$N ratios are respectively {$270\pm 50$} and
{$290\pm 50$} for HCN and HNC. These values are in remarkable agreement
with the direct determinations. The total weighted average including
both direct and indirect measures is $270\pm 30$. All the
$^{14}$N/$^{15}$N ratios derived in OMC-2 FIR4 are plotted in
Fig.~\ref{fig:ratiocompared}.

{As discussed for each species, the narrow emission of HCN, HNC and CN, from which the $^{14}$N/$^{15}$N ratio is measured, shows a very low excitation temperature ($6-9$ K). Without large scale emission maps for these species, we cannot firmly establish that they trace extended parent gas of the protocluster but it seems a reasonable interpretation. HC$_3$N narrow emission traces at least 2 components: a relatively cold gas (T$_{ex}$ = 25K), that our recent observations obtained with the 30m telescope show to be extended, and a warmer component (T$_{ex}$ $\sim$ 50 K). A more sophisticated analysis of the HC$_3$N emission, included also the broad emission will be presented in a forthcoming article (Jaber Al-Edhari et al., in prep.). In addition, we also tentatively measured with the HCN isotopologues the $^{14}$N/$^{15}$N ratio in the broad emission, detected for the main isotopologue of all the species studied here, except N$_2$H$^+$. This ratio appears perfectly compatible with the value derived from the narrow emission. Our data suggest that this broad emission is warmer than the narrow one but do not allow to estimate the emission size. We hope that the interferometric (ALMA and NOEMA) data that we will soon obtain towards this source will allow to understand the nature of this emission.}

\subsection{Comparison with other galactic sources}
\begin{figure*}
  \centering
\plotone{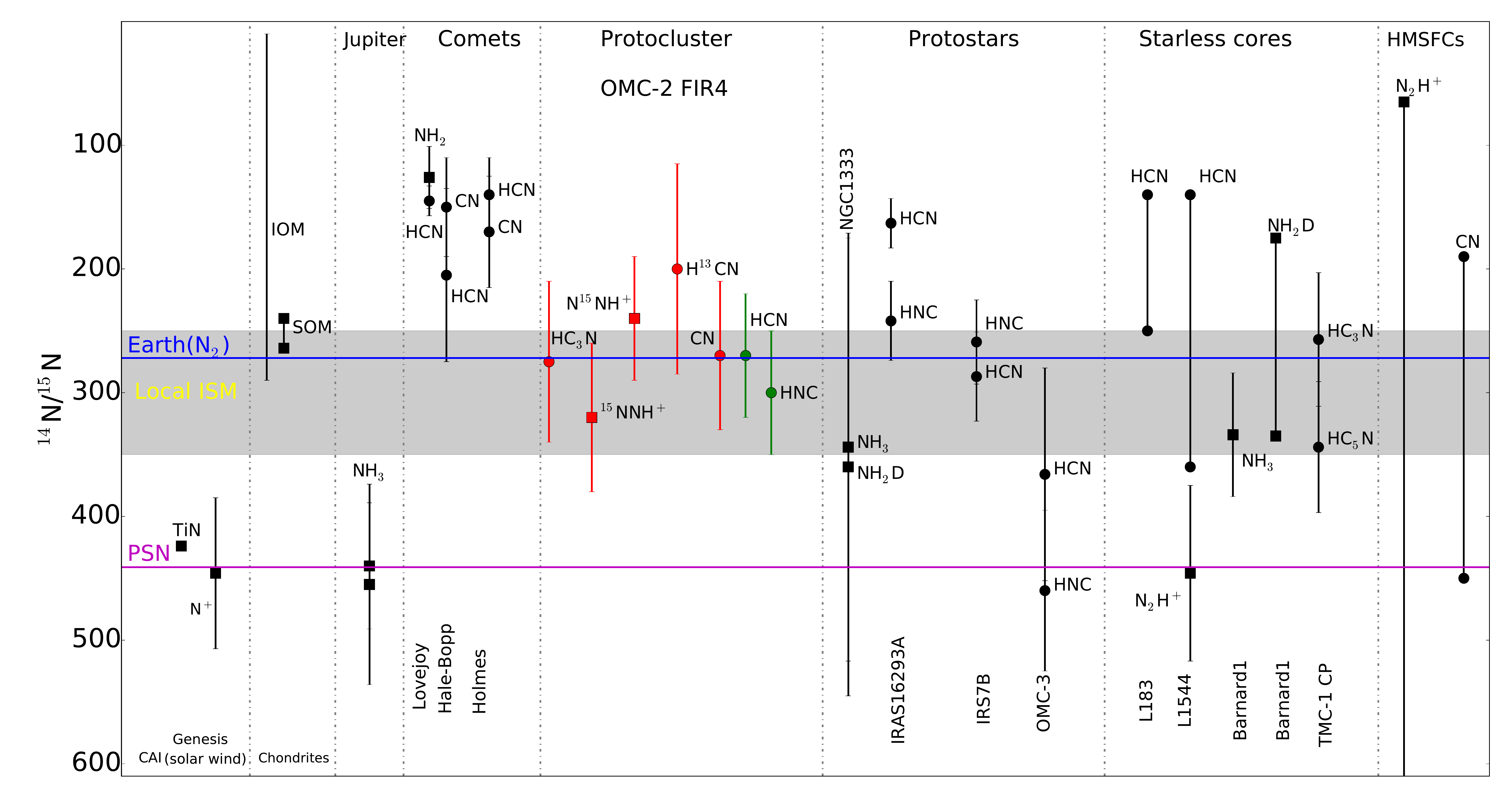}
  \caption{Nitrogen isotopic composition of Solar System objects
    compared with different sources (figure adapted
    from~\cite{Hily-Blant:2013}). The blue horizontal line represents
    the nitrogen isotopic composition of the terrestrial
    atmosphere($^{14}$N/$^{15}$N=272), while the magenta horizontal
    line indicates the protosolar nebula (PSN) value of $441\pm 6$~\cite{Marty:2012}. Square and circle symbols correspond to
    amine and nitrile functional groups, respectively. IOM means
    Insoluble Organic Matter, SOM Soluble Organic Matter, and CAI
    Calcium-, Aluminum-rich Inclusions. Red and green symbols
    represent respectively direct and indirect measurements of
    $^{14}$N/$^{15}$N in OMC-2 FIR4, assuming $^{12}$C/$^{13}$C
    ={$50\pm 5$} (this work). Protostars data are from
    ~\cite{Wampfler:2014}, starless cores data
    from~\cite{Hily-Blant:2013} and \cite{Taniguchi:2017}, and high
    mass star forming cores data from ~\cite{Fontani:2015} (where
    HMSC: high-mass starless core). {The local galactic ratio range, shown as a grey zone, takes into account the measurements reported in ~\cite{Adande:2012} and ~\cite{Hily-Blant:2017}}   
}    
  \label{fig:ratiocompared}
\end{figure*}

Measurements of the the $^{14} $N/$^{15}$N ratio in starless cores
(e.g.~\cite{Hily-Blant:2013}) and in protostars
(e.g.~\cite{Wampfler:2014}) seem to indicate (see Fig.~\ref{fig:ratiocompared}) that the ratios derived
from molecules carrying the amine functional group (NH$_3$,
N$_2$H$^+$) are larger than the ratios derived from molecules carrying
the nitrile functional group (CN, HCN), a chemical origin being
proposed for this effect. However, none of the studied sources shows a
complete set of measurements from different tracers so that it is very
difficult to distinguish between variations from source to source and
from molecule to molecule. On the other hand, our results, which rely
on a set of five different species that trace the same (cold extended)
gas, and that belong to the nitrile and amine families, do not show
any significant difference. In contrast, they are very similar, and
they remarkably agree with the present local $^{14}$N/$^{15}$N
galactic ratio of $\sim 300$ as derived from observations
(~\cite{Adande:2012};{ ~\cite{Hily-Blant:2017} }and references therein) and predicted by models of galactic CNO evolution (e.g. ~\cite{Romano:2017}).

%%%%%%%%%% N-Ratio %%%%%%%%
%%%%%%%%%%%%%%%%%%%%%%%%%%%%%%%%%%
%
%

Our observations show that in OMC-2 FIR4, which is the
best analogue of the Sun progenitor, there is no $^{15}$N
fractionation, compared to the nowadays value at the same (8 kpc) galactic
center distance. This is in agreement with the recent model
predictions by~\cite{Roueff:2015}. 

{In conclusion, the presented measurements of $^{14}$N/$^{15}$N seem to be at odd with the previous measurements in pre-stellar cores and protostars (see Fig.~\ref{fig:ratiocompared} ), which, depending on the used species, suggest $^{15}$N enrichment or deficiency with respect to the local ISM value. It is possible that this discrepancy is due to the different spatial scale probed by our and the others’ observations. Specifically, while the cold and large scale gas might not be $^{15}$N enriched, local smaller scale clumps might present this enrichment (or deficiency). On going interferometric observations towards OMC-2 FIR4 will verify this possibility. If this is the case, the enrichment of the Solar System bodies could find an explanation in the ISM chemistry. If, on the contrary, the new observations would confirm the absence of $^{15}$N enrichment also at small scales, then the $^{15}$N enrichment in Solar System bodies must have another nature.}
%
%
%\newpage
%%%%%%%%%%%%%%  Table Opacity checks : suppressed %%%%%%%%%%%%

\section*{Acknowledgements}

We acknowledge the financial support from the university of
Al-Muthanna and the ministry of higher education and scientific
research in Iraq. We acknowledge the funding from the European Research Council (ERC), project DOC (the Dawn of Organic Chemistry), contract number 741002.
We warmly thank Pierre Hily-Blant for fruitful
discussions.

\software{CASSIS \citep{Vastel:2015}, GILDAS}

%%%%%%%%%%%%%  Opacity checks %%%%%%%%%%%%
\newpage
\appendix
\section{Observed Lines}
%%%% Fig. A1 HC3N %%%% 
\begin{figure}[htb]
  \centering
 \plotone{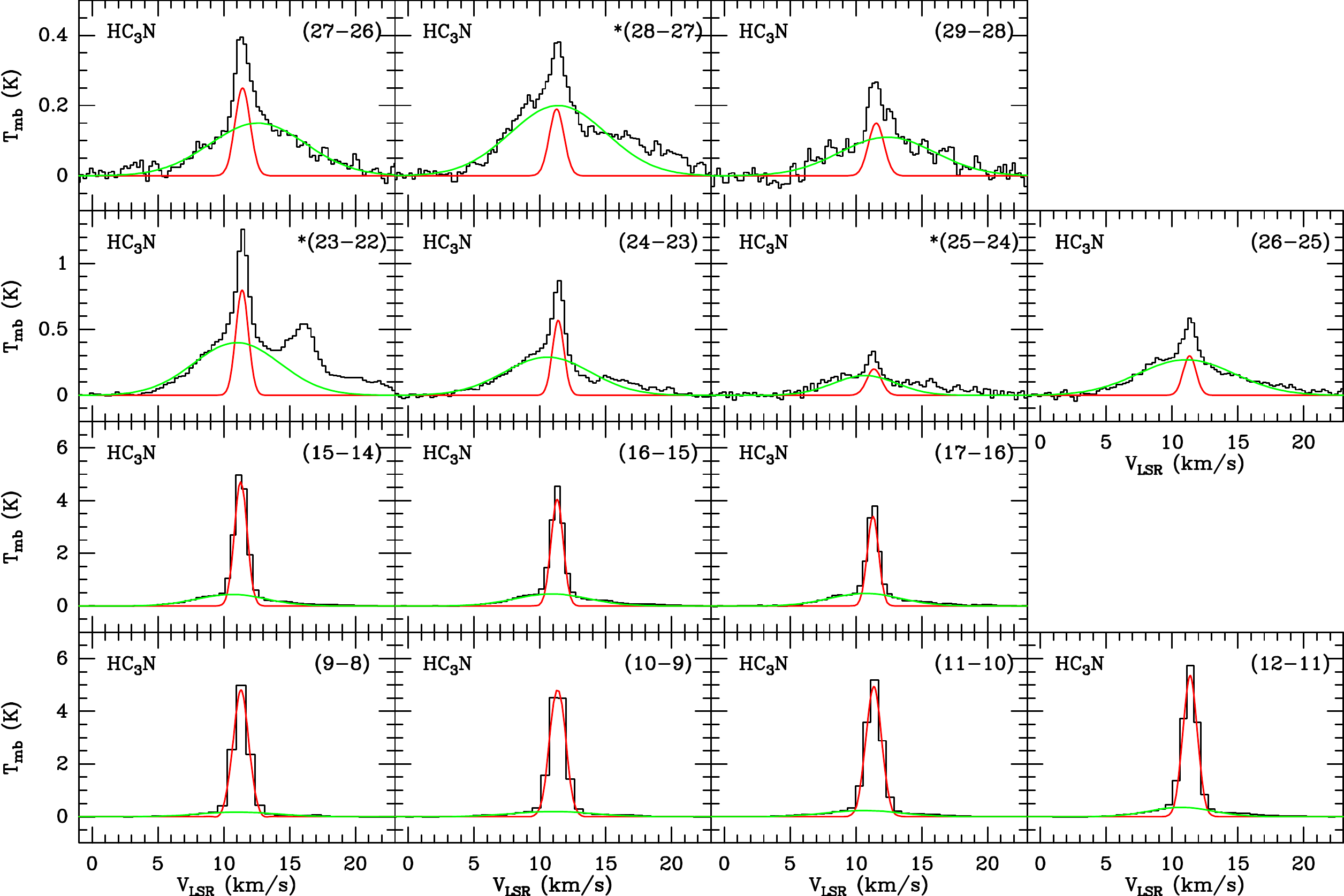}
  \caption{HC$_3$N observed spectra
    and two components Gaussian fits to the lines. The temperature
    scale is main beam temperature. Due to significant overlaping with other emissions or calibration problem, the lines indicated by an asterisc (*) have not been included in our analysis nor reported in Table~\ref{tab:HC3N}.}
  \label{fig:obs_HC3N}
\end{figure}

%%%% Fig. A2 H13CCCN %%%% 
\begin{figure}[htb]
  \centering
 \plotone{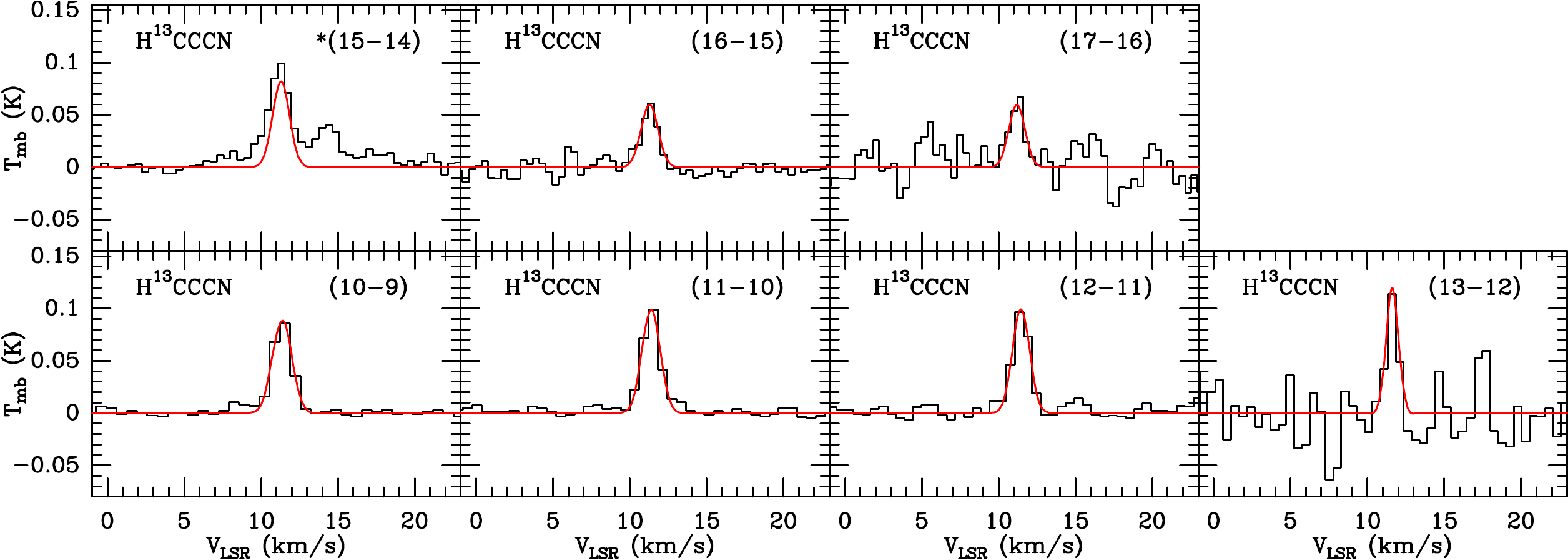}
  \caption{ H$^{13}$CCCN  observed spectra
   and Gaussian fit to the lines. The temperature
    scale is main beam temperature. Due to significant overlaping with other emissions the line indicated by an asterisc (*) has not been included in our analysis nor reported in Table~\ref{tab:HC3N}.}
  \label{fig:obs_H13CCCN}
\end{figure}

%%%% Fig. A3 HC13CCN %%%% 
\begin{figure}[htb]
  \centering
 \plotone{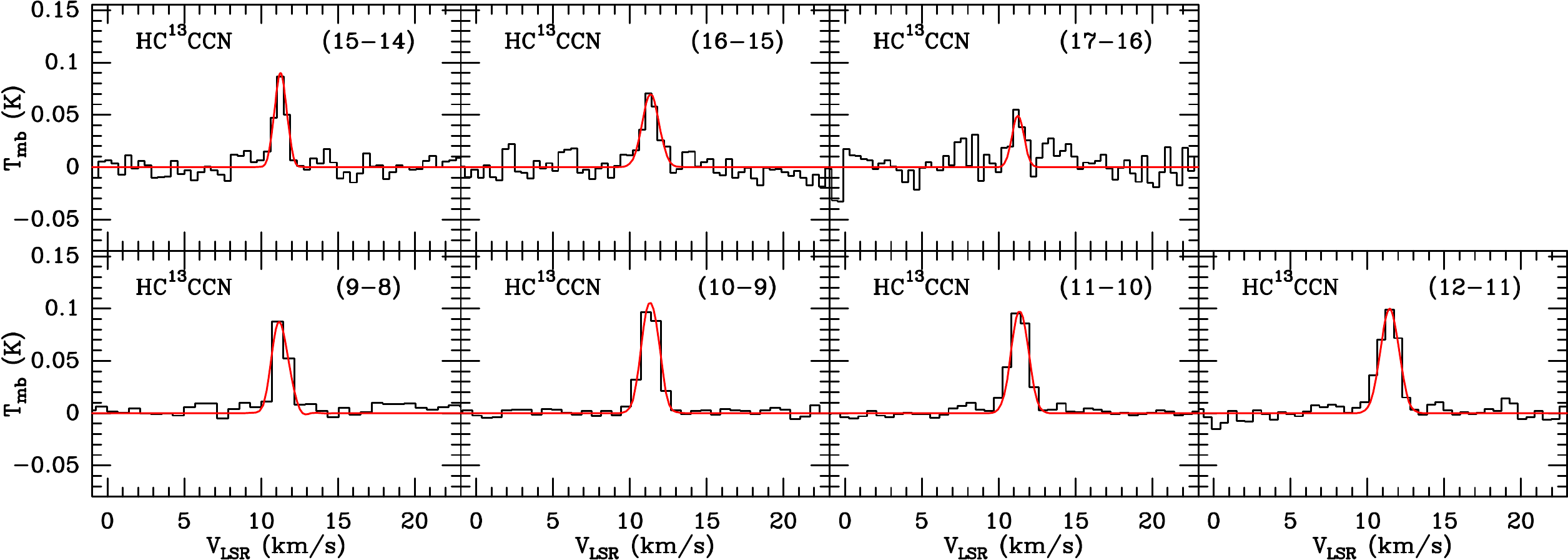}
  \caption{HC$^{13}$CCN observed spectra
    and Gaussian fit to the lines. The temperature
    scale is main beam temperature.}
  \label{fig:obs_HC13CCN}
\end{figure}

%%%% Fig. A4 HCC13CN %%%% 
\begin{figure}[htb]
  \centering
 \plotone{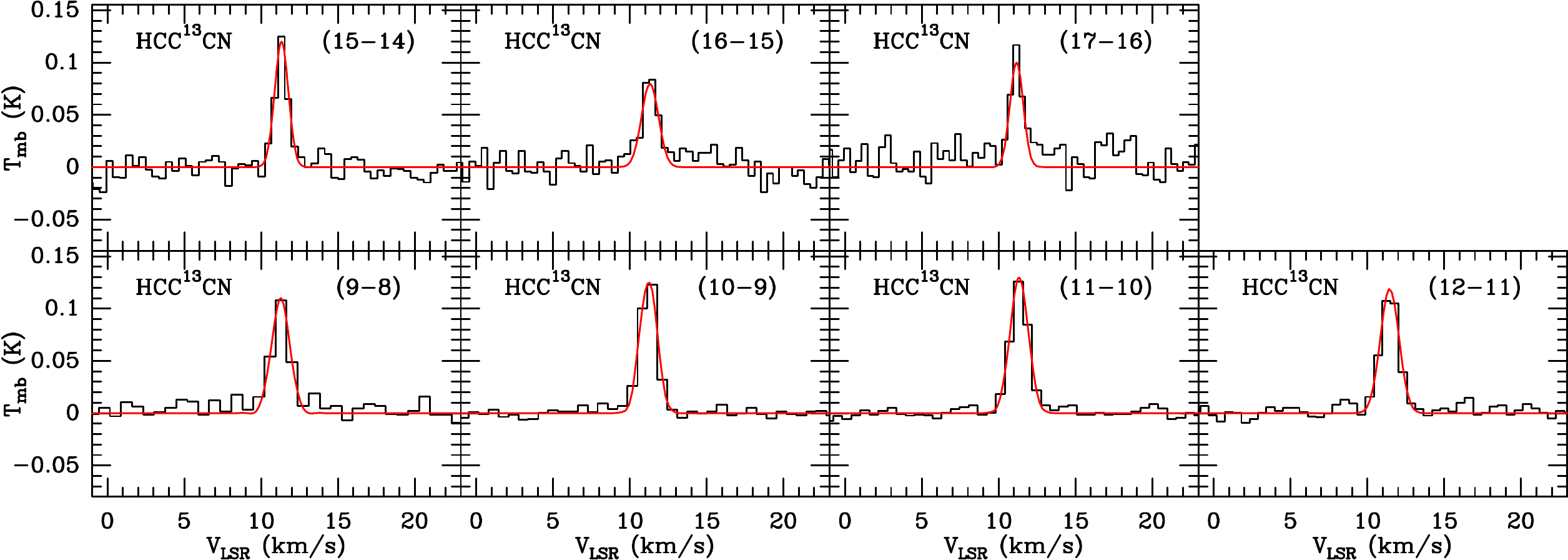}
  \caption{HCC$^{13}$CN observed spectra 
    and Gaussian fit to the lines. The temperature
    scale is main beam temperature.}
  \label{fig:obs_HCC13CN}
\end{figure}

%%%% Fig. A5 HCCC15N %%%% 
\begin{figure}[htb]
  \centering
 \plotone{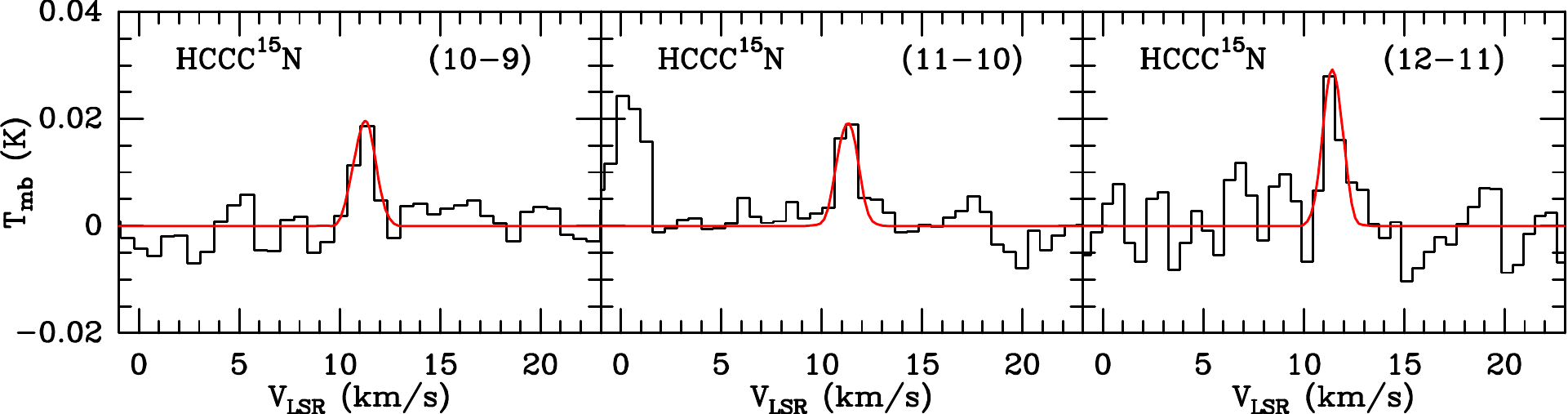}
  \caption{HCCC$^{15}$N observed spectra 
    and Gaussian fit to the lines. The temperature
    scale is main beam temperature.}
  \label{fig:obs_HCCC15N}
\end{figure}

%%%% Fig. A6 HCN %%%% 
\begin{figure}[htb]
  \centering
 \plotone{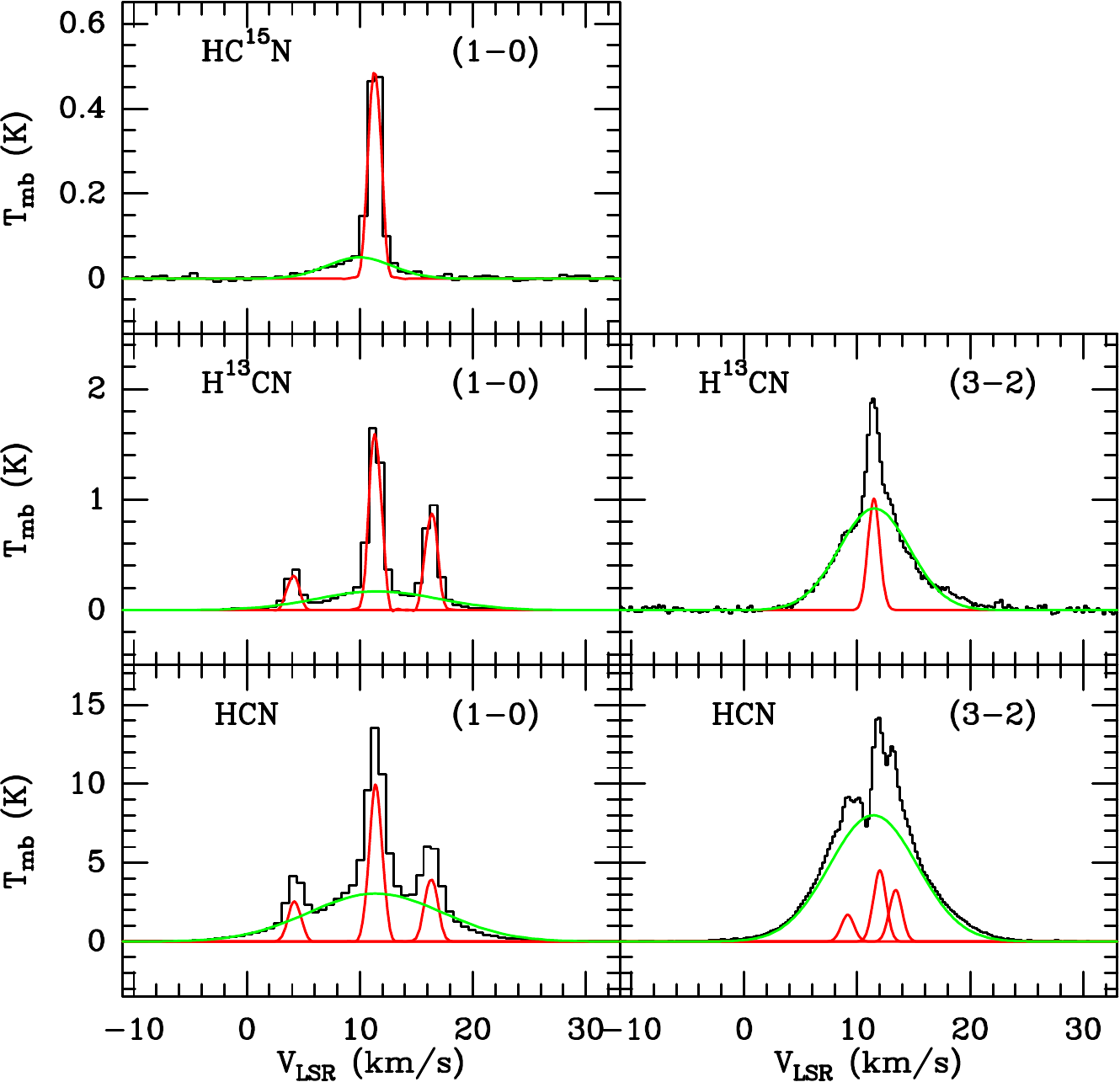}
  \caption{HCN 
    and its isotopologues observed spectra. Two components Gaussian fits to the lines are superimposed to the spectra. The temperature
    scale is main beam temperature.}
  \label{fig:obs_HCN}
\end{figure}

%%%% Fig. A7-8 HNC & N2H+ %%%% 
\begin{figure}[htb]
  \centering
 \plotone{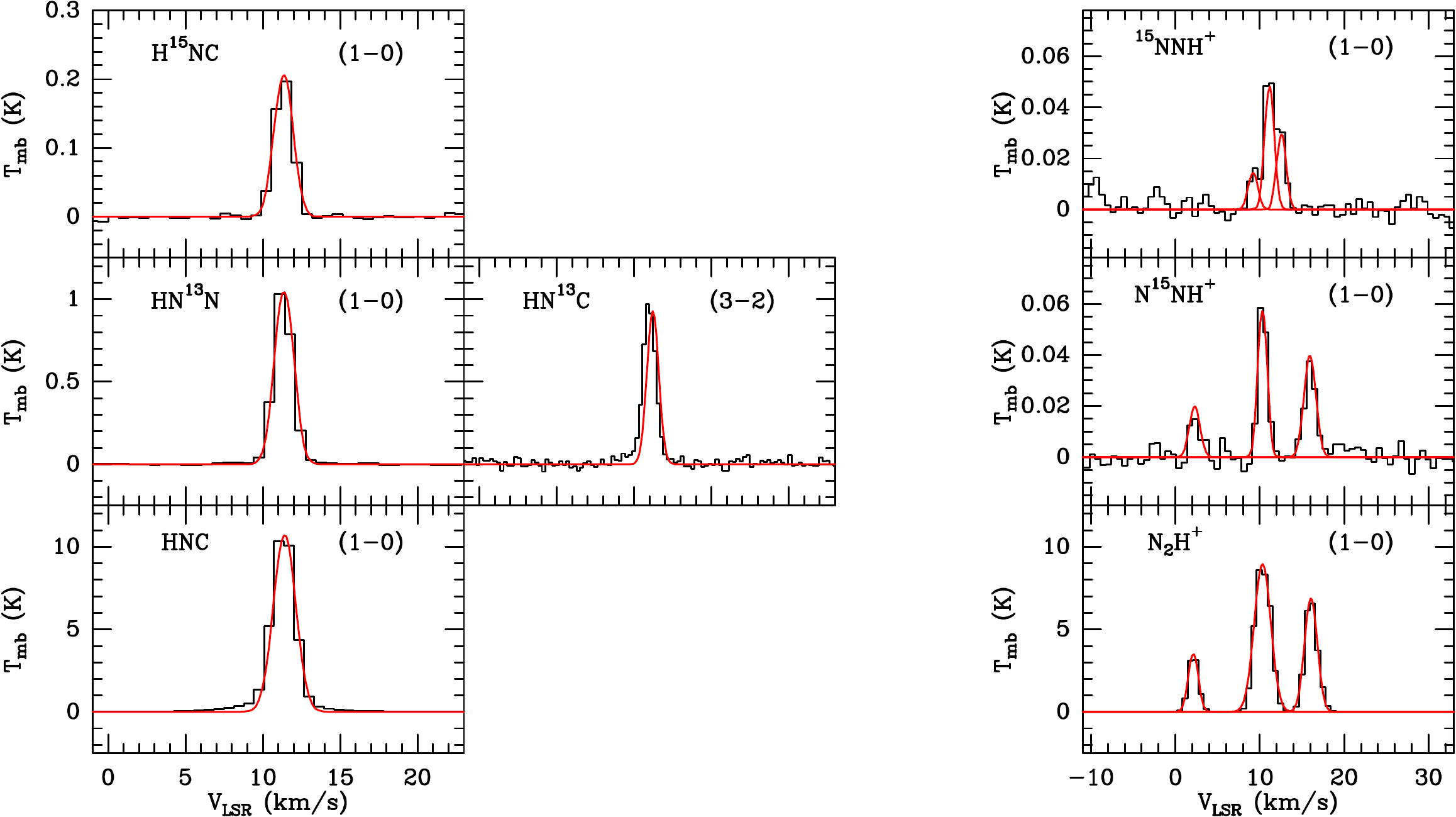}
  \caption{HNC, N$_2$H$^+$  and their isotopologues observed spectra.
     Gaussian fits to the lines are superimposed to the spectra. The temperature     scale is main beam temperature.}
  \label{fig:obs_HNC-N2H+}
\end{figure}

%%%% Fig. A9 CN %%%% 
\begin{figure}[htb]
  \centering
 \plotone{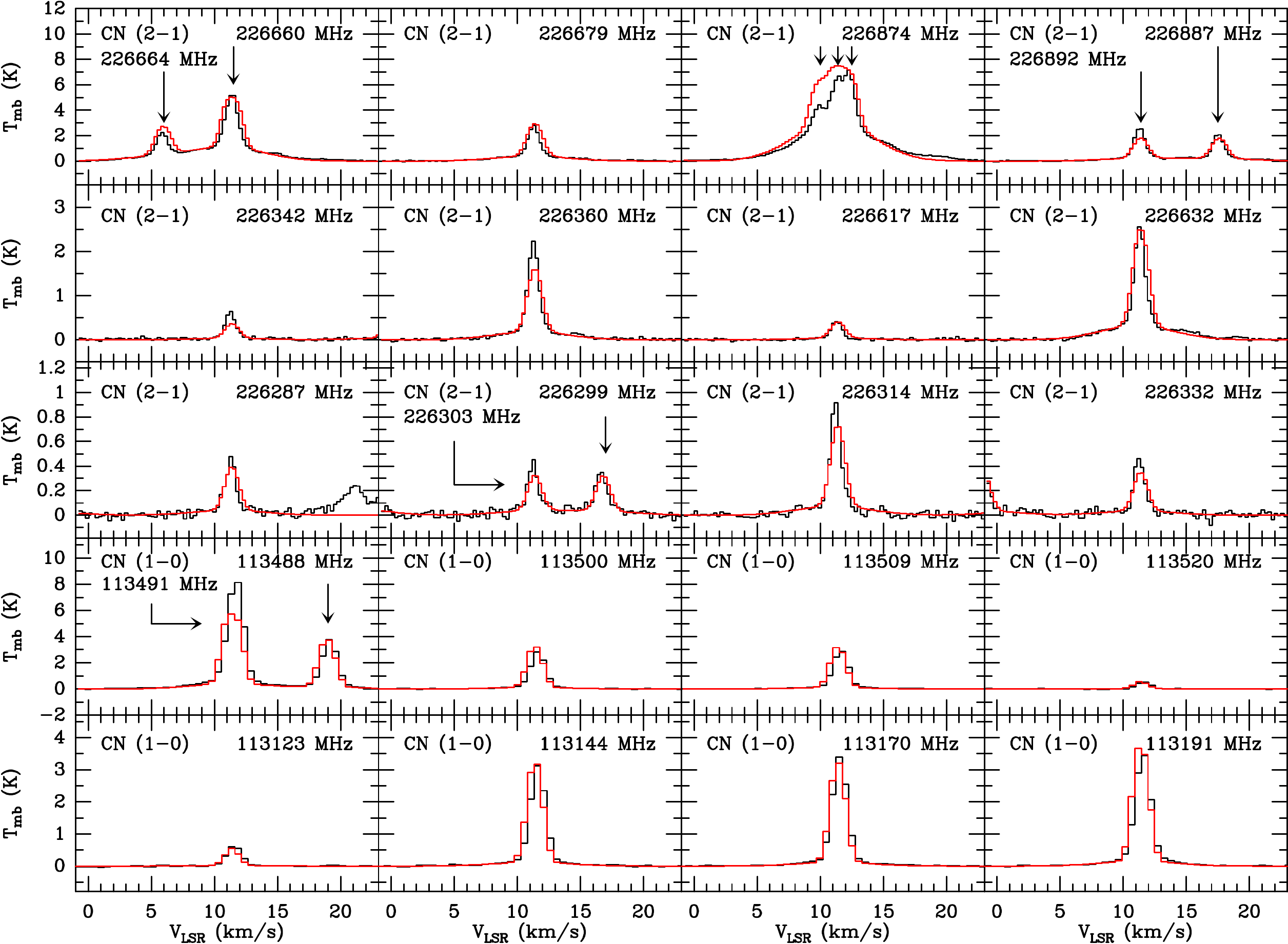}
  \caption{CN observed spectra. The profiles superimposed to the spectra have been calculated with a two components LTE modeling (see text). When several close hyperfine components are present, the frequency of the main one is given. The temperature scale is main beam temperature.}
  \label{fig:obs_CN}
\end{figure}

%%%% Fig. A10 CN %%%% 
\begin{figure}[htb]
  \centering
 \plotone{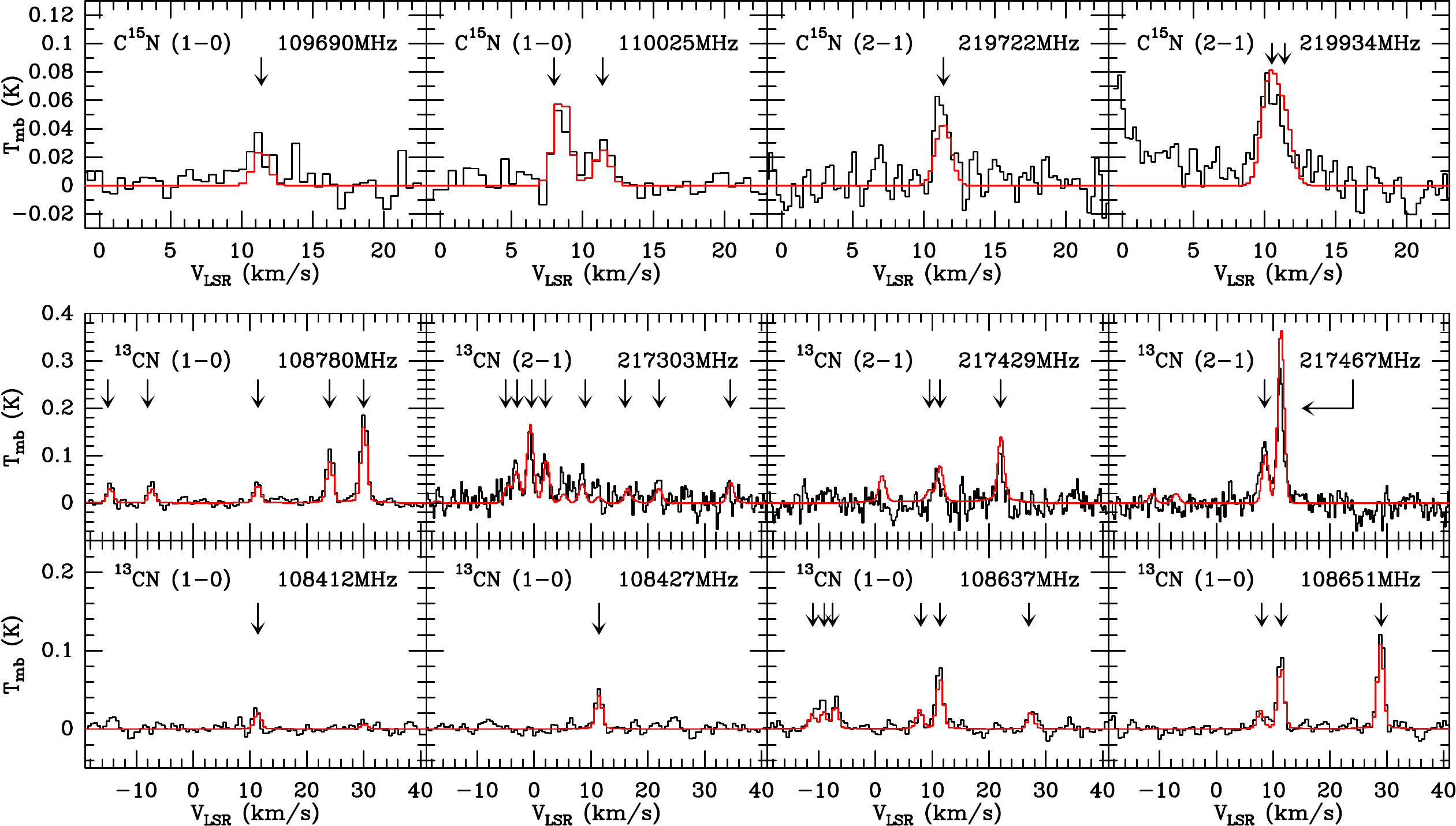}
  \caption{$^{13}$CN and C$^{15}$Nobserved spectra. The profiles superimposed to the spectra have been calculated with a LTE modeling (see text). When several close hyperfine components are present, the frequency of the main one is given. The temperature scale is main beam temperature.}
  \label{fig:obs_13C15N}
\end{figure}

\section{Opacity checks of hyperfine components}
\begin{ThreePartTable}
\begin{center}
\begin{longtable}{|c|c|c|c|c|c|c|c|c|c|}
\caption{Relative fluxes of the hyperfine components.}
\label{tab:check}\\
\hline 
Species   &    Transition       & Frequency  & E$_{up}$ &  g$_{up}$ & A$_{ij}$   & R\_theo.$^{(1)}$  & $\int$T$_{mb}$dv$^{(2)}$ & R\_obs.$^{(3)}$ \\
          &                     & [MHz]      & [K]      &           & [s$^{-1}$] &                  & [K.km.s$^{-1}$]    &           \\
           \hline 
\endfirsthead
\multicolumn{10}{c}%
{{series \tablename\ \thetable{} -- continued from previous page}} \\
\hline
 Species   &    Transition       & Frequency  & E$_{up}$ &  g$_{up}$ & A$_{ij}$   & R\_theo.$^{(1)}$  & $\int$T$_{mb}$dv$^{(2)}$ & R\_obs.$^{(3)}$ \\
          &                     & [MHz]      & [K]      &           & [S$^{-1}$] &                  & [K.km.s$^{-1}$]     &                      \\
           \hline
\endhead

\hline \multicolumn{10}{|r|}{{Continued on next page}} \\ \hline
\endfoot

\hline 
\endlastfoot

HCN        & 1$_1$ $-$ 0$_1$        & 88630.42    & 4.25     & 3      & 2.43$\times 10^{-5}$     & 3.0           & 8.4(0.9)         & 1.6(0.2)                \\
           & 1$_2$ $-$ 0$_1$        & 88631.85    & 4.25     & 5      & 2.43$\times 10^{-5}$     & 5.0           & 13.5(2.0)        & 2.6(0.5)               \\
           & 1$_0$ $-$ 0$_1$*       & 88633.94    & 4.25     & 1      & 2.43$\times 10^{-5}$     & 1.0           & 5.2(0.5)         & 1.0(0.1)                    \\ 
         \hline
H$^{13}$CN & 1$_1$ $-$ 0$_1$       & 86338.77     & 4.14     & 3      & 2.22$\times 10^{-5}$     & 3.0           & 1.3(0.1)         & 2.9(0.5)                     \\
           & 1$_2$ $-$ 0$_1$       & 86340.18     & 4.14     & 5      & 2.22$\times 10^{-5}$     & 5.0           & 2.2(0.2)         & 5.1(0.9)                     \\  
           & 1$_0$ $-$ 0$_1$*      & 86342.27     & 4.14     & 1      & 2.22$\times 10^{-5}$     & 1.0           & 0.44(0.06)       & 1.0(0.2)                       \\
          \hline
NNH$^+$    & 1$_1$ $-$ 0$_1$       & 93171.88     & 4.47     & 9      & 3.63$\times 10^{-5}$     & 3.0           & 13.1(1.5)        & 2.5(0.4)                \\  
           & 1$_2$ $-$ 0$_1$       & 93173.70     & 4.47     & 15     & 3.63$\times 10^{-5}$     & 5.0           & 21.3(2.5)        & 4.1(0.6)               \\
           & 1$_0$ $-$ 0$_1$*      & 93176.13     & 4.47     & 3      & 3.63$\times 10^{-5}$     & 1.0           & 5.3(0.5)         & 1.0(0.1)                \\
          \hline
$^{15}$NNH$^+$ & 1$_2$ $-$ 0$_1$   & 90263.91     & 4.38     & 3      & 3.30$\times 10^{-5}$     & 3.0           & 0.036(0.005)     & 2.4(0.7)                  \\           
               & 1$_1$ $-$ 0$_1$   & 90263.49     & 4.38     & 5      & 3.30$\times 10^{-5}$     & 5.0           & 0.08(0.01)       & 5.0(1.5)                \\ 
               & 1$_0$ $-$ 0$_1$*  & 90264.50     & 4.38     & 1      & 3.30$\times 10^{-5}$     & 1.0           & 0.015(0.004)     & 1.0(0.4)                 \\ 
           \hline
N$^{15}$NH$^+$ & 1$_1$ $-$ 0$_1$   & 91204.26     & 4.33     & 3      & 3.40$\times 10^{-5}$     & 3.0           & 0.059(0.008)     & 2.4(0.6)              \\ 
               & 1$_2$ $-$ 0$_1$   & 91205.99     & 4.33     & 5      & 3.40$\times 10^{-5}$     & 5.0           & 0.09(0.02)       & 3.8(1.2)                \\ 
               & 1$_0$ $-$ 0$_1$*  & 91208.52     & 4.33     & 1      & 3.40$\times 10^{-5}$     & 1.0           & 0.025(0.006)     & 1.0(0.3)                  \\           
           \hline
CN    & 1$_{0~1/2~1/2}$ $-$ 0$_{0~1/2~1/2}$* & 113123.37 & 5.43   & 2 & 1.29$\times 10^{-6}$    & 1.0            & 0.9(0.1)         & 1.1(0.2)              \\
      & 1$_{0~1/2~1/2}$ $-$ 0$_{0~1/2~3/2}$  & 113144.19 & 5.43   & 2 & 1.05$\times 10^{-5}$    & 8.1            & 4.9(0.5)         & 5.8(0.8)             \\
      & 1$_{0~1/2~3/2}$ $-$ 0$_{0~1/2~1/2}$  & 113170.54 & 5.43   & 4 & 5.14$\times 10^{-6}$    & 7.9            & 5.3(0.5)         & 6.2(0.9)                \\
      & 1$_{0~1/2~3/2}$ $-$ 0$_{0~1/2~3/2}$  & 113191.33 & 5.43   & 4 & 6.68$\times 10^{-6}$    & 10.3           & 5.6(0.6)         & 6.6(1.0)              \\
      & 1$_{0~3/2~3/2}$ $-$ 0$_{0~1/2~1/2}$  & 113488.14 & 5.45   & 4 & 6.73$\times 10^{-6}$    & 10.4           & 6.0(0.7)         & 7.0(1.1)               \\
      & 1$_{0~3/2~5/2}$ $-$ 0$_{0~1/2~3/2}$  & 113490.99 & 5.45   & 6 & 1.19$\times 10^{-5}$    & 27.5           & 14.0(1.6)        & 16.4(2.5)             \\
      & 1$_{0~3/2~1/2}$ $-$ 0$_{0~1/2~1/2}$  & 113499.64 & 5.45   & 2 & 1.06$\times 10^{-5}$    & 8.2            & 4.3(0.4)         & 5.1(0.7)                \\
      & 1$_{0~3/2~3/2}$ $-$ 0$_{0~1/2~3/2}$  & 113508.93 & 5.45   & 4 & 5.19$\times 10^{-6}$    & 8.0            & 4.5(0.5)         & 5.3(0.8)              \\
      & 1$_{0~3/2~1/2}$ $-$ 0$_{0~1/2~3/2}$* & 113520.42 & 5.45   & 2 & 1.30$\times 10^{-6}$    & 1.0            & 0.76(0.08)       & 0.9(0.1)               \\
& 2$_{0~3/2~1/2}$ $-$ 1$_{0~3/2~1/2}$ & 226287.43 & 16.31 &2	&	1.03$\times 10^{-5}$	&	4.6	&	 0.5(0.1) & 5.1(1.1) \\
 & 2$_{0~3/2~1/2}$ $-$ 1$_{0~3/2~3/2}$ & 226298.92 & 16.31 &2	&	8.23$\times 10^{-6}$	&	3.7	&	 0.6(0.2) & 5.8(1.7) \\
 & 2$_{0~3/2~3/2}$ $-$ 1$_{0~3/2~1/2}$ & 226303.08 &  16.31 &4	&	4.17$\times 10^{-6}$	&	3.7	&	 0.5(0.1) & 4.9(1.2) \\
 & 2$_{0~3/2~3/2}$ $-$ 1$_{0~3/2~3/2}$ & 226314.54 & 16.31 &4	&	9.91$\times 10^{-6}$	&	8.8	&	1.0(0.2) &10.1(2.1) \\
 & 2$_{0~3/2~3/2}$ $-$ 1$_{0~1/2~5/2}$ & 226332.54 & 16.31 &4	&	4.55$\times 10^{-6}$	&	4.0	&	 0.5(0.1) &5.0	(1.1) \\
 & 2$_{0~3/2~5/2}$ $-$ 1$_{0~3/2~3/2}$ & 226341.93 & 16.31 &6	&	3.16$\times 10^{-6}$	&	4.2	&	0.7(0.1) &6.6(1.4)\\
 & 2$_{0~3/2~5/2}$ $-$ 1$_{0~3/2~5/2}$ & 226359.87 & 16.31 &6	&	1.61$\times 10^{-5}$	&	21.4	&	 2.6(0.6) &25.4(5.3)\\
 & 2$_{0~3/2~1/2}$ $-$ 1$_{0~1/2~3/2}$ & 226616.56 & 16.31 &2	&	1.07$\times 10^{-5}$	&	4.8	&	 0.5(0.1) & 4.6(0.9) \\
 & 2$_{0~3/2~3/2}$ $-$ 1$_{0~1/2~3/2}$ & 226632.19 & 16.31 &4	&	4.26$\times 10^{-5}$	&	37.9	&	3.2(0.7) & 31.2(6.8)\\
 & 2$_{0~3/2~5/2}$ $-$ 1$_{0~1/2~3/2}$ & 226659.58 & 16.31 &6	&	9.47$\times 10^{-5}$	&	126.2	&	8.3(1.7) & 80(17)\\
 & 2$_{0~3/2~1/2}$ $-$ 1$_{0~1/2~1/2}$ & 226663.70 & 16.31 &2	&	8.46$\times 10^{-5}$	&	37.6	&	 3.2(0.7) &30.9(6.6) \\
 & 2$_{0~3/2~3/2}$ $-$ 1$_{0~1/2~1/2}$ & 226679.38 & 16.31 &4	&	5.27$\times 10^{-5}$	&	46.8	&	 3.6(0.8) & 35.0	(7.3)\\
 & 2$_{0~5/2~5/2}$ $-$ 1$_{0~3/2~5/2}$ & 226892.12 & 16.34 &6	&	1.81$\times 10^{-5}$	&	24.1	&	3.2(0.7) & 30.9	(6.5)\\
 & 2$_{0~5/2~3/2}$ $-$ 1$_{0~3/2~5/2}$* & 226905.38 & 16.34 &4  &	1.13$\times 10^{-6}$	&	1.0	&	 0.10(0.03) & 1.0(0.3)\\      

    \hline
$^{13}$CN  & 1$_{1/2~0~1}$ $-$ 0$_{1/2~1~1}$ & 108412.86 & 5.23   & 3 & 3.1$\times 10^{-6}$     & 1.1            & 0.03(0.01)       & 1.1(0.5)             \\
           & 1$_{1/2~0~1}$ $-$ 0$_{1/2~1~2}$ & 108426.89 & 5.23   & 3 & 6.3$\times 10^{-6}$     & 2.3            & 0.069(0.008)     & 2.1(0.7)               \\
           & 1$_{1/2~1~0}$ $-$ 0$_{1/2~0~1}$ & 108631.12 & 5.21   & 1 & 9.6$\times 10^{-6}$     & 1.2            & 0.04(0.02)       & 1.1(0.7)               \\
           & 1$_{1/2~1~1}$ $-$ 0$_{1/2~0~1}$ & 108636.92 & 5.21   & 3 & 9.6$\times 10^{-6}$     & 3.5            & 0.11(0.02)       & 3.4(1.1)               \\
           & 1$_{1/2~1~2}$ $-$ 0$_{1/2~0~1}$ & 108651.30 & 5.21   & 5 & 9.8$\times 10^{-6}$     & 5.9            & 0.18(0.03)       & 5.4(1.9)               \\
           & 1$_{3/2~1~2}$ $-$ 0$_{1/2~1~2}$ & 108657.65 & 5.24   & 5 & 7.2$\times 10^{-6}$     & 4.4            & 0.13(0.02)       & 4.0(1.3)             \\
           & 1$_{3/2~2~3}$ $-$ 0$_{1/2~1~2}$ & 108780.20 & 5.25   & 7 & 1.1$\times 10^{-5}$     & 8.9            & 0.27(0.05)       & 8.2(2.8)            \\
           & 1$_{3/2~2~2}$ $-$ 0$_{1/2~1~1}$ & 108782.37 & 5.25   & 5 & 7.8$\times 10^{-6}$     & 4.7            & 0.16(0.03)       & 4.9(1.6)             \\
           & 1$_{3/2~2~1}$ $-$ 0$_{1/2~1~0}$ & 108786.98 & 5.25   & 3 & 5.7$\times 10^{-6}$     & 2.1            & 0.06(0.01)       & 2.0(0.7)            \\
           & 1$_{3/2~2~1}$ $-$ 0$_{1/2~1~1}$ & 108793.75 & 5.25   & 3 & 4.5$\times 10^{-6}$     & 1.6            & 0.06(0.01)       & 1.9(0.6)             \\
           & 1$_{3/2~2~2}$ $-$ 0$_{1/2~1~2}$ & 108796.40 & 5.25   & 5 & 2.8$\times 10^{-6}$     & 1.7            & 0.05(0.02)       & 1.7(0.8)              \\
           & 1$_{3/2~1~2}$ $-$ 0$_{1/2~1~1}$ & 108643.59 & 5.24   & 5 & 2.6$\times 10^{-6}$     & 1.6            & 0.06(0.01)       & 1.7(0.6)               \\
           & 1$_{3/2~1~0}$ $-$ 0$_{1/2~1~1}$ & 108644.35 & 5.24   & 1 & 9.6$\times 10^{-6}$     & 1.2            & 0.06(0.02)       & 1.8(0.7)           \\
           & 1$_{3/2~1~1}$ $-$ 0$_{1/2~1~1}$*& 108645.06 & 5.24   & 3 & 2.7$\times 10^{-6}$     & 1.0            & 0.03(0.01)       & 1.0(0.4)             \\
           & 1$_{3/2~1~1}$ $-$ 0$_{1/2~1~2}$ & 108658.95 & 5.24   & 3 & 3.3$\times 10^{-6}$     & 1.2            & 0.03(0.01)       & 1.0(0.5)               \\
           & 2$_{5/2~3~3}$ $-$ 1$_{3/2~2~2}$ & 217467.15 & 15.69 & 7	&	8.92$\times 10^{-5}$ 	&	0.6	&	0.36(0.08) &	0.6(0.1)  \\
           & 2$_{5/2~3~2}$ $-$ 1$_{3/2~2~1}$ & 217469.15 & 15.69 & 5	&	8.43$\times 10^{-5}$	&	0.4	& 0.23(0.05) &	0.4(0.1)  \\
        \hline
C$^{15}$N  & 1$_{1/2~1}$ $-$ 0$_{1/2~1}$*   & 109689.61 & 5.27    & 3 & 7.10$\times 10^{-6}$    & 1.0            & 0.05(0.02)       & 1.0(0.5)             \\
           & 1$_{3/2~1}$ $-$ 0$_{1/2~1}$    & 110023.54 & 5.28    & 3 & 7.16$\times 10^{-6}$    & 1.0            & 0.05(0.02)       & 1.0(0.4)               \\
           & 1$_{3/2~2}$ $-$ 0$_{1/2~1}$    & 110024.59 & 5.28    & 5 & 1.09$\times 10^{-5}$    & 2.6            & 0.08(0.02)       & 1.7(0.7)               \\
           & 2$_{3/2~2}$ $-$ 1$_{1/2~1}$ & 219722.49 & 15.81 & 5	&	8.67$\times 10^{-5}$	&	0.3 & 0.08(0.02) &	0.4(0.1) \\
           & 2$_{5/2~2}$ $-$ 1$_{3/2~1}$ & 219934.04 & 15.84 & 5	&	9.36$\times 10^{-5}$	&	0.7 & 0.10(0.03) & 0.6(0.1)\\     
\end{longtable}
  \begin{tablenotes}
    \item[*]{\footnotesize Weakest detected line of the hyperfine
        structure. This is the reference to compute R$_{theo}$ and R$_{obs}$}
    \item[1]{\footnotesize Ratio of the line G$_{up}$ A$_{ij}$ product relative to the reference line; equal to the fluxes ratio in LTE optically thin conditions, neglecting the frequency differences.}
    \item[2]{\footnotesize Observed lines fluxes derived from gaussian fits ; the errors include fit and calibration uncertainties.}
    \item[3]{\footnotesize Ratio of observed fluxes relative to the reference line. For CN $(N=1-0)$, the reference flux is averaged over the 2 reference lines.}
    \end{tablenotes}
\end{center}
\end{ThreePartTable}
%%%%%%%%%%%%%%%%%%%%%%%%%%%%%%%%%%%%

%%%%%%%%%%%%%%%%%%%% REFERENCES %%%%%%%%%%%%%%%%%%

%\bibliographystyle{aasjournal}
%\bibliography{omc2ratioj} % if your bibtex file is called example.bib

\end{document}